\documentclass[11pt]{article}

\usepackage{latexsym,amssymb,fullpage}
\usepackage{amsmath,amscd}
\usepackage{amsmath,amsfonts,graphics,graphicx,epsfig,bbm}
\usepackage{epstopdf}
\usepackage{amsmath,amscd,latexsym,amssymb,color}
\usepackage[all]{xy}\CompileMatrices

\newtheorem{prop}{Proposition}\def\PRO{\begin{prop}}\def\ORP{\end{prop}}
\newtheorem{coro}{Corollary}\def\COR{\begin{coro}}\def\ROC{\end{coro}}
\newtheorem{theo}{Theorem}\def\TH{\begin{theo}}\def\HT{\end{theo}}
\newtheorem{defi}[prop]{Definition}\def\DE{\begin{defi}}\def\ED{\end{defi}}
\newtheorem{lemme}[prop]{Lemma}\def\LE{\begin{lemme}}\def\EL{\end{lemme}}
\newcommand{\AR}[2][c]{$$\begin{array}[#1]{lllllllllllllll}#2\end{array}$$}
\newtheorem{algo}{Algorithm}

\def\sem#1{[\hspace{-.35ex}[#1]\hspace{-.35ex}]}
\def\mybox#1{\framebox{$#1$}}
\def\tup#1{\langle#1\rangle}
\def\tuple#1{\langle#1\rangle}

\def\qed{$\Box$}
\def\emptyset{\varnothing}
\def\setminus{\smallsetminus}

\def\ens#1{\{#1\}}
\def\ie{\textit{i.e.}}
\def\eg{\textit{e.g.}}
\def\aka{\textit{a.k.a.}}

\def\st{^\star}
\def\al{\alpha}
\def\ba{\beta} 
\def\ga{\gamma}
\let\da\delta
\def\eps{\epsilon}

\def\la{\lambda} 

\def\ta{\theta}

\def\Ga{\Gamma}

\def\rar{\rightarrow}

\def\Rar{\Rightarrow}

\def\lrar{\longrightarrow}

\def\slar#1{\stackrel{#1}{\lrar}}

\def\mcl{\mathcal} 
\def\mbb{\mathbb} 
 
\def\mfr{\mathcal}
\def\MA#1{\left(\begin{matrix}#1\end{matrix}\right)}
\def\EQ#1{\begin{eqnarray}#1\end{eqnarray}}
\def\AL{\begin{algo}}\def\LA{\end{algo}}

\def\iN{\iota}
\def\oU{o}
\def\ei#1{e^{i#1}}
\def\emi#1{e^{{-i}#1}}
\def\eta{\ta}

\def\ad{^\dag}

\def\hil#1{\mathfrak H_{#1}}

\def\bhil#1{{\mathfrak B}_{#1}}

\def\ket#1{{|}#1\rangle}
\def\bra#1{\langle#1{|}}
\def\oqb#1{\ket{\hskip-.1ex+_{#1}}}

\def\oqbn#1{\ket{\hskip-.1ex-_{#1}}}
\def\oqbb#1{\bra{\hskip-.1ex+_{#1}}}
\def\oqbnb#1{\bra{\hskip-.1ex-_{#1}}}

\def\ctR{\mathop{\wedge}\hskip-.4ex} 
\def\ctwo{{\mbb C}^2}

\def\ztwo{{\mbb Z}_2}
\def\ost{\frac1{\sqrt2}}

\def\pit{\frac\pi2}
\def\pif{\frac\pi4}
\def\norm#1{\|#1\|}

\def\Rr#1#2{J_{#1}^{#2}}
\def\Cx#1{\cx{#1}{}}
\def\Cz#1{\cz{#1}{}}
\def\cz#1#2{Z_{#1}^{#2}}
\def\cx#1#2{X_{#1}^{#2}}
\def\G{J}

\def\MS#1#2#3#4{{}^{#4}[{M}_{#2}^{#1}]^{#3}}
\def\ms#1#2#3{\MS{#1}{#2}{#3}{}}
\def\Ms#1#2{{M}_{#2}^{#1}}
\def\M#1#2{{M}_{#2}^{#1}}

\def\et#1#2{E_{#1#2}}
\def\nq#1{N_{#1}}
\def\Nq#1#2{N^{#2}_{#1}}
\def\Cp#1#2{Z^{#2}_{#1}}
\def\cp#1#2#3{Z_{#1}^{#2,#3}}
\def\CO#1{A_{#1}}
\def\ss#1#2{S_{#1}^{#2}}

\def\TMLR#1#2#3#4#5#6#7#8{{}^{(#7,#8)}[{M}_{#3#4}^{#1,#2}]^{(#5,#6)}}
\def\TMR#1#2#3#4#5#6{[{M}_{#3#4}^{#1,#2}]^{(#5,#6)}}
\def\TML#1#2#3#4#5#6{{}^{(#5,#6)}[{M}_{#3#4}^{#1,#2}]}
\def\tM#1#2#3#4{M_{#3#4}^{#1,#2}}
\def\mLR#1#2#3#4{{}^{#4}[{M}_{#2}^{#1}]^{#3}}
\def\mR#1#2#3{\mLR{#1}{#2}{#3}{}}
\def\mL#1#2#3{\mLR{#1}{#2}{}{#3}}
\def\m#1#2{{M}_{#2}^{#1}}

\title{The Measurement Calculus}
\author{Vincent Danos\\Universit\'e Paris~7 \& CNRS\\
  {\small\texttt{vincent.danos@pps.jussieu.fr}}
  \and
  Elham Kashefi\\ IQC - University of Waterloo\\ Christ Church - Oxford\\
  {\small\texttt{elham.kashefi@comlab.ox.ac.uk}}
  \and
  Prakash Panangaden\\ McGill University\\
  {\small\texttt{prakash@cs.mcgill.ca}}}

\begin{document}
\date{\empty}
\maketitle

\begin{abstract}
  
  Measurement-based quantum computation has emerged from the physics
  community as a new approach to quantum computation where the notion of
  measurement is the main driving force of computation.  This is in
  contrast with the more traditional circuit model which is based on
  unitary operations.  Among measurement-based quantum computation methods,
  the recently introduced one-way quantum computer~\cite{RB01} stands out
  as fundamental.
  
  We develop a rigorous mathematical model underlying the one-way quantum
  computer and present a concrete syntax and operational semantics for
  programs, which we call patterns, and an algebra of these patterns
  derived from a denotational semantics.  More importantly, we present a
  calculus for reasoning locally and compositionally about these patterns.
  We present a rewrite theory and prove a general standardization theorem
  which allows all patterns to be put in a semantically equivalent standard
  form.  Standardization has far-reaching consequences: a new physical
  architecture based on performing all the entanglement in the beginning,
  parallelization by exposing the dependency structure of measurements and
  expressiveness theorems.
  
  Furthermore we formalize several other measurement-based models \eg
  \,Teleportation, Phase and Pauli models and present compositional
  embeddings of them into and from the one-way model.  This allows us to
  transfer all the theory we develop for the one-way model to these models.
  This shows that the framework we have developed has a general impact on
  measurement-based computation and is not just particular to the one-way
  quantum computer.

\end{abstract}

%%%%%%%%%%%%%%%%%%%%%%%%%%%%%%%

\section{Introduction}

The emergence of quantum computation has changed our perspective on many
fundamental aspects of computing: the nature of information and how it
flows, new algorithmic design strategies and complexity classes and the very
structure of computational models \cite{NC00}.  New challenges have been
raised in the physical implementation of quantum computers.  This paper is
a contribution to a nascent discipline: quantum programming languages.

This is more than a search for convenient notation, it is an investigation
into the structure, scope and limits of quantum computation.  The main
issues are questions about how quantum processes are defined, how quantum
algorithms compose, how quantum resources are used and how classical and
quantum information interact.  

Quantum computation emerged in the early 1980s with Feynman's observations
about the difficulty of simulating quantum systems on a classical computer.
This hinted at the possibility of turning around the issue and exploiting
the power of quantum systems to perform computational tasks more
efficiently than was classically possible.  In the mid 1980s
Deutsch~\cite{Deutsch87} and later Deutsch and Jozsa~\cite{DJ92} showed how
to use superposition -- the ability to produce linear combinations of
quantum states -- to obtain computational speedup.  This led to interest in
algorithm design and the complexity aspects of quantum computation by
computer scientists.  The most dramatic results were Shor's celebrated
polytime factorization algorithm~\cite{Shor94} and Grover's sublinear
search algorithm~\cite{Grover98a}.  Remarkably one of the problematic
aspects of quantum theory, the presence of non-local correlation -- an
example of which is called ``entanglement'' -- turned out to be crucial for
these algorithmic developments.

If efficient factorization is indeed possible in practice, then much of
cryptography becomes insecure as it is based on the difficulty of
factorization.  However, entanglement makes it possible to design
unconditionally secure key distribution~\cite{BB84,Ekert91}.  Furthermore,
entanglement led to the remarkable -- but simple -- protocol for transferring
quantum states using only classical communication~\cite{teleport93}; this
is the famous so-called ``teleportation'' protocol.  There continues to be
tremendous activity in quantum cryptography, algorithmic design, complexity
and information theory.  Parallel to all this work there has been intense
interest from the physics community to explore possible implementations,
see, for example, \cite{NC00} for a textbook account of some of these
ideas.

On the other hand, only recently has there been significant interest in
quantum programming languages; i.e.\ the development of formal syntax and
semantics and the use of standard machinery for reasoning about quantum
information processing.  The first quantum programming languages were
variations on imperative probabilistic languages and emphasized logic and
program development based on weakest preconditions~\cite{Sanders00,Omer01}.
The first definitive treatment of a quantum programming language was the
flowchart language of Selinger~\cite{Selinger04}.  It was based on
combining classical control, as traditionally seen in flowcharts, with
quantum data.  It also gave a denotational semantics based on completely
positive linear maps.  The notion of quantum weakest preconditions was
developed in~\cite{DP04}.  Later people proposed languages based on
quantum control~\cite{AG05}.  The search for a sensible notion of
higher-type computation~\cite{SV05,vanTonder04} continues, but is
problematic~\cite{selinger04b}.  

A related recent development is the work of Abramsky and
Coecke~\cite{AbrCoe04,Coe04} where they develop a categorical
axiomatization of quantum mechanics.  This can be used to verify the
correctness of quantum communication protocols.  It is very interesting
from a foundational point of view and allows one to explore exactly what
mathematical ingredients are required to carry out certain quantum
protocols.  This has also led to work on a categorical quantum
logic~\cite{AD04}.  

The study of quantum communication protocols has led to formalizations
based on process algebras~\cite{GR04,JL04} and to proposals to use model
checking for verifying quantum protocols.  A survey and a complete list of
references on this subject up to 2005 is available~\cite{Gay05}.

These ideas have proven to be of great utility in the world of classical
computation.  The use of logics, type systems, operational semantics,
denotational semantics and semantic-based inference mechanisms have led to
notable advances such as: the use of model checking for verification,
reasoning compositionally about security protocols, refinement-based
programming methodology and flow analysis.

The present paper applies this paradigm to a very recent development:
measurement-based quantum computation.  None of the cited research on
quantum programming languages is aimed at measurement-based computation.
On the other hand, the work in the physics literature does not clearly
separate the conceptual layers of the subject from implementation issues.
A formal treatment is necessary to analyze the foundations of
measurement-based computation.

So far the main framework to explore quantum computation has been the
circuit model~\cite{Deutsch89}, based on unitary evolution.  This is very
useful for algorithmic development and complexity analysis \cite{BV97}.
There are other models such as quantum Turing machines~\cite{Deutsch85} and
quantum cellular automata~\cite{Watrous95,vanDam96,DS96,SW04}.  Although
they are all proved to be equivalent from the point of view of expressive
power, there is no agreement on what is the canonical model for exposing
the key aspects of quantum computation.

Recently physicists have introduced novel ideas based on the use of
measurement and entanglement to perform computation
\cite{Gott99,RB01,mqqcs,Nielsen03}.  This is very different from the
circuit model where measurement is done only at the end to extract
classical output.  In measurement-based computation the main operation to
manipulate information and control computation is measurement.  This is
surprising because measurement creates indeterminacy, yet it is used to
express deterministic computation defined by a unitary evolution.

The idea of computing based on measurements emerged from the teleportation
protocol~\cite{teleport93}.  The goal of this protocol is for an agent to
transmit an unknown qubit to a remote agent without actually sending the
qubit.  This protocol works by having the two parties share a maximally
entangled state called a Bell pair.  The parties perform \emph{local}
operations -- measurements and unitaries -- and communicate only classical
bits.  Remarkably, from this classical information the second party can
reconstruct the unknown quantum state.  In fact one can actually use this
to compute via teleportation by choosing an appropriate
measurement~\cite{Gott99}.  This is the key idea of measurement-based
computation.

It turns out that the above method of computing is actually universal.
This was first shown by Gottesman and Chuang~\cite{Gott99} who used
two-qubit measurements and given Bell pairs.  Later
Nielsen~\cite{Nielsen03} showed that one could do this with only 4-qubit
measurements with no prior Bell pairs, however this works only
probabilistically.  Leung~\cite{L03} improved this to two qubits, but her
method also works only probabilistically.  Later Perdrix and
Jorrand~\cite{Perdrix04,PJ04} gave the minimal set measurements to perform
universal quantum computing -- but still in the probabilistic setting --
and introduced the state-transfer and measurement-based quantum Turing
machine.  Finally the one-way computer was invented by Raussendorf and
Briegel~\cite{RB01,RB02} which used only single-qubit measurements with a
particular multi-party entangled state, the cluster state.

More precisely, a computation consists of a phase in which a collection of
qubits are set up in a standard entangled state.  Then measurements are
applied to individual qubits and the outcomes of the measurements may be
used to determine further measurements.  Finally -- again depending on
measurement outcomes -- local unitary operators, called corrections, are
applied to some qubits; this allows the elimination of the indeterminacy
introduced by measurements.  The phrase ``one-way'' is used to emphasize
that the computation is driven by irreversible measurements.

There are at least two reasons to take measurement-based models seriously:
one conceptual and one pragmatic.  The main pragmatic reason is that the
\emph{one-way} model is believed by physicists to lend itself to easier
implementations~\cite{Nielsen04,CMJ04,BR04,TPKV04,TPKV05,nature05,KPA05,BES05,CCWD05,BBFM05}.
Physicists have investigated various properties of the cluster state and
have accrued evidence that the physical implementation is scalable and
robust against decoherence
\cite{Schl03,graphstates,DAB03,VDD04b,VDD04,MP04,GHW05,HDB05,DHN06}.
Conceptually the measurement-based model highlights the role of
entanglement and separates the quantum and classical aspects of
computation; thus it clarifies, in particular, the interplay between
classical control and the quantum evolution process.

Our approach to understanding the structural features of measurement-based
computation is to develop a formal calculus.  One can think of this as an
``assembly language'' for measurement-based computation.  Ours is the first
programming framework specifically based on the one-way model.  We first
develop a notation for such classically correlated sequences of
entanglements, measurements, and local corrections.  Computations are
organized in patterns\footnote{We use the word ``pattern'' rather than
  ``program'', because this corresponds to the
  commonly used terminology in the physics literature.}, and we give a
careful treatment of the composition and tensor product (parallel
composition) of patterns.  We show next that such pattern combinations
reflect the corresponding combinations of unitary operators.  An easy proof
of universality follows.

So far, this is primarily a clarification of what was already known from
the series of papers introducing and investigating the properties of the
one-way model~\cite{RB01,RB02,mqqcs}.  However, we work here with an
extended notion of pattern, where inputs and outputs may overlap in any way
one wants them to, and this results in more efficient -- in the sense of
using fewer qubits -- implementations of unitaries.  Specifically, our
universal set consists of patterns using only 2 qubits. From it
we obtain a 3 qubit realization of the $R_z$ rotations and 
a 14 qubit realization for the controlled-$U$ family: a significant
reduction over the hitherto known implementations.

The main point of this paper is to introduce a calculus of local equations
over patterns that exploits some special algebraic properties of the
entanglement, measurement and correction operators.  More precisely, we use
the fact that that 1-qubit $XY$ measurements are closed under conjugation
by Pauli operators and the entanglement command belongs to the normalizer
of the Pauli group; these terms are explained in the appendix.  We show
that this calculus is sound in that it preserves the interpretation of
patterns.  Most importantly, we derive from it a simple algorithm by which
any general pattern can be put into a standard form where entanglement is
done first, then measurements, then corrections.  We call this
\emph{standardization}.

The consequences of the existence of such a procedure are far-reaching.
Since entangling comes first, one can prepare the entire entangled state
needed during the computation right at the start: one never has to do ``on
the fly'' entanglements.  Furthermore, the rewriting of a pattern to
standard form reveals parallelism in the pattern computation.  In a general
pattern, one is forced to compute sequentially and to strictly obey the
command sequence, whereas, after standardization, the dependency structure
is relaxed, resulting in lower computational depth complexity.  Last, the
existence of a standard form for any pattern also has interesting
corollaries beyond implementation and complexity matters, as it follows
from it that patterns using no dependencies, or using only the restricted
class of Pauli measurements, can only realize a unitary belonging to the
Clifford group, and hence can be efficiently simulated by a classical
computer \cite{Gottesman97}.

As we have noted before, there are other methods for measurement-based
quantum computing: the teleportation technique based on two-qubit
measurements and the state-transfer approach based on single qubit
measurements and incomplete two-qubit measurements.  We will analyze the
teleportation model and its relation to the one-way model.  We will show
how our calculus can be smoothly extended to cover this case as well as new
models that we introduce in this paper.  We get several benefits from our
treatment.  We get a workable syntax for handling the dependencies of
operators on previous measurement outcomes just by mimicking the one
obtained in the one-way model.  This has never been done before for the
teleportation model.  Furthermore, we can use this embedding to obtain a
standardization procedure for the models.  Finally these extended calculi
can be compositionally embedded back in the original one-way model.  This
clarifies the relation between different measurement-based models and shows
that the one-way model of Raussendorf and Briegel is the canonical one.

This paper develops the one-way model \emph{ab initio} but certain concepts
that the reader may be unfamiliar with: qubits, unitaries, measurements,
Pauli operators and the Clifford group are in an appendix.  These are also
readily accessible through the very thorough book of Nielsen and
Chuang~\cite{NC00}.

In the next section we define the basic model, followed by its operational
and denotational semantics, for completeness a simple proof of universality
is given in section~\ref{s:univ}, this has appeared earlier in the physics
literature~\cite{generator04}, in section~\ref{s-cal} we develop the rewrite
theory and prove the fundamental standardization theorem.  In
section~\ref{s:examples} we develop several examples that illustrate the use
of our calculus in designing efficient patterns.  In section~\ref{s:expr} we
prove some theorems about the expressive power of the calculus in the
absence of adaptive measurements.  In section~\ref{s:other} we discuss other
measurement-based models and their compositional embedding to and from the
one-way model.  In section~\ref{s:conc} we discuss further directions and
some more related work.  In the appendix we review basic notions of quantum
mechanics and quantum computation.

%%%%%%%%%%%%%%%%%%%%%%%%%%%%%%%

\section{Measurement Patterns}
We first develop a notation for 1-qubit measurement based computations.  The
basic commands one can use in a pattern are: 
\begin{itemize}
\item 1-qubit auxiliary preparation $\nq i$
\item 2-qubit entanglement operators $\et ij$
\item 1-qubit measurements $\M{\al}i$
\item and 1-qubit Pauli operators corrections $\Cx i$ and $\Cz i$
\end{itemize}

The indices $i$, $j$ represent the qubits on which each of these operations
apply, and $\al$ is a parameter in $[0,2\pi]$.  Expressions involving angles
are always evaluated modulo $2\pi$.  These types of command will be referred
to as $N$, $E$, $M$ and $C$.  Sequences of such commands, together with two
distinguished -- possibly overlapping -- sets of qubits corresponding to
inputs and outputs, will be called \emph{measurement patterns}, or simply
patterns.  These patterns can be combined by composition and tensor product.

Importantly, corrections and measurements are allowed to depend on previous
measurement outcomes.  We shall prove later that patterns without these
classical dependencies can only realize unitaries that are in the Clifford
group.  Thus, dependencies are crucial if one wants to define a universal
computing model; that is to say, a model where all unitaries over
$\otimes^n\ctwo$ can be realized.  It is also crucial to develop a
notation that will handle these dependencies. This is what we do now.

\subsection{Commands}
Preparation $\nq i$ prepares qubit $i$ in state $\ket{+}_i$.  The
entanglement commands are defined as $\et ij:=\ctR Z_{ij}$
(controlled-$Z$), while the correction commands are the Pauli operators
$\Cx i$ and $\Cz i$.

Measurement $\M\al i$ is defined by orthogonal projections on
\AR{
\oqb\al&:=&\ost(\ket0+ e^{i\al}\ket1)\\
\oqbn\al&:=&\ost(\ket0-\ei\al\ket1)
}

followed by a trace-out operator.  The parameter $\al \in [0,2\pi]$ is
called the \emph{angle} of the measurement.  For $\al=0$, $\al=\pit$, one
obtains the $X$ and $Y$ Pauli measurements.  Operationally, measurements
will be understood as destructive measurements, consuming their qubit.  The
\emph{outcome} of a measurement done at qubit $i$ will be denoted by
$s_i\in\ztwo$.  Since one only deals here with patterns where qubits are
measured at most once (see condition (D1) below), this is unambiguous.  We
take the specific convention that $s_i=0$ if under the corresponding
measurement the state collapses to $\oqb\al$, and $s_i=1$ if to $\oqbn\al$.

Outcomes can be summed together resulting in expressions of the form
$s=\sum_{i\in I} s_i$ which we call \emph{signals}, and where the summation
is understood as being done in $\ztwo$.  We define the \emph{domain} of a
signal as the set of qubits on which it depends.

As we have said before, both corrections and measurements may depend on
signals.  Dependent corrections will be written $\cx is$ and $\cz is$ and
dependent measurements will be written $\MS \al ist $, where $s,t \in
\ztwo$ and $\al\in [0,2\pi]$.  The meaning of dependencies for corrections
is straightforward: $\cx i0=\cz i0=I$, no correction is applied, while
$\cx i1=\Cx i$ and $\cz i1=\Cz i$.  In the case of dependent measurements,
the measurement angle will depend on $s$, $t$ and $\al$ as follows: \EQ{
\label{msem}
\MS{\al} ist&:=&\M{(-1)^s\al+t\pi} i
}
so that, depending on the parities of $s$ and $t$, one may have to modify
the $\al$ to one of $-\al$, $\al+\pi$ and $-\al+\pi$.  These modifications
correspond to conjugations of measurements under $X$ and $Z$: \EQ {
  X_i\M{{\al}}i X_i&=&\M{{-\al}}i\label{xmx}\\
  Z_i\M{{\al}}i Z_i&=&\M{{\al+\pi}}i\label{zmz} } accordingly, we will refer to
them as the $X$ and $Z$-actions.  Note that these two actions commute,
since $-\al+\pi=-\al-\pi$ up to $2\pi$, and hence the order in which one
applies them does not matter.

As we will see later, relations (\ref{xmx}) and (\ref{zmz}) are key to the
propagation of dependent corrections, and to obtaining patterns in the
standard entanglement, measurement and correction form.  Since the measurements
considered here are destructive, the above equations actually simplify
to 
\EQ{
\label{e:simp1}
\M{{\al}}i X_i&=&\M{{-\al}}i\\
\label{e:simp2}
\M{{\al}}i Z_i&=&\M{{\al-\pi}}i
}
Another point worth noticing is that the domain of the signals of a
dependent command, be it a measurement or a correction, represents the set
of measurements which one has to do before one can determine the actual
value of the command.  

We have completed our catalog of basic commands, including dependent ones,
and we turn now to the definition of measurement patterns.  For convenient
reference, the language syntax is summarized in Figure ~\ref{syntax}.

\begin{figure}[t]
\begin{center}
$$
\begin{array}{lllr}
S&:=& 0,\,1,\,s_i,\,S+S&\quad\hbox{Signals}\\
A &:=&N_i&\quad\hbox{Preparations}\\
&& \et ij&\quad\hbox{Entanglements}\\
&&\MS\al ist&\quad\hbox{Measurements}\\
&&\cx is,\,\cz is&\quad\hbox{Corrections}
\end{array}
$$
\caption{1-qubit based measurement language syntax}
\label{syntax}
\end{center}
\end{figure}

\subsection{Patterns}
\DE
Patterns consists of three finite sets $V$, $I$, $O$, together with two
injective maps $\iN:I\rar V$ and $\oU:O\rar V$ 
and a finite sequence of commands $A_n\ldots A_1$, read from right to left,
applying to qubits in $V$ in that order, i.e.\ $A_1$  first and $A_n$ last,
such that: 
\begin{description}
\item[(D0)] no command depends on an outcome not yet measured;
\item[(D1)] no command acts on a qubit already measured;
\item[(D2)] no command acts on a qubit not yet prepared, unless it is an
  input qubit; 
\item[(D3)] a qubit $i$ is measured if and only if $i$ is not an output.
\end{description} 
\ED

The set $V$ is called the pattern \emph{computation space}, and we write
$\hil V$ for the associated quantum state space $\otimes_{i\in V}\ctwo$.  To
ease notation, we will omit the maps $\iN$ and $\oU$,
and write simply $I$, $O$ instead of $\iN(I)$ and $\oU(O)$.  Note, however,
that these maps are useful to define classical manipulations of the quantum
states, such as permutations of the qubits.  The sets $I$, $O$ are
called respectively the pattern \emph{inputs} and \emph{outputs}, and we
write $\hil I$, and $\hil O$ for the associated quantum state spaces.
The sequence $A_n\ldots A_1$ is called the pattern \emph{command
  sequence}, while the triple $(V,I,O)$ is called the pattern
\emph{type}.

To run a pattern, one prepares the input qubits in some input state
$\psi\in\hil I$, while the non-input qubits are all set to the $\ket{+}$
state, then the commands are executed in sequence, and finally the result
of the pattern computation is read back from outputs as some $\phi\in\hil
O$.  Clearly, for this procedure to succeed, we had to impose the (D0),
(D1), (D2) and (D3) conditions.  Indeed if (D0) fails, then at some point
of the computation, one will want to execute a command which depends on
outcomes that are not known yet.  Likewise, if (D1) fails, one will try to
apply a command on a qubit that has been consumed by a measurement (recall
that we use destructive measurements).  Similarly, if (D2) fails, one will
try to apply a command on a non-existent qubit.  Condition (D3) is there to
make sure that the final state belongs to the output space $\hil O$, \ie,
that all non-output qubits, and only non-output qubits, will have been
consumed by a measurement when the computation ends.

We write (D) for the conjunction of our definiteness conditions (D0),
(D1), (D2) and (D3).  Whether a given pattern satisfies (D) or not is
statically verifiable on the pattern command sequence.  We could have
imposed a simple type system to enforce these constraints but, in the
interests of notational simplicity, we chose not to do so.  

Here is a concrete example:
\AR{
\mfr H&:=&
(\ens{1,2},\ens{1},\ens{2},\cx 2{s_1}\M01\et12\nq 2)
}
with computation space $\ens{1,2}$, inputs $\ens{1}$, and outputs
$\ens{2}$.  To run $\mfr H$, one first prepares the first qubit in some
input state $\psi$, and the second qubit in state $\ket+$, then these are
entangled to obtain $\ctR Z_{12}(\psi_1\otimes\ket+_2)$.  Once this is
done, the first qubit is measured in the $\ket+$, $\ket-$ basis.  Finally an
$X$ correction is applied on the output qubit, if the measurement outcome
was $s_1=1$.  We will do this calculation in detail later, and prove that
this pattern implements the Hadamard operator $H$.

In general, a given pattern may use auxiliary qubits that are neither
input nor output qubits.  Usually one tries to use as few such qubits as
possible, since these contribute to the \emph{space complexity} of the
computation.

A last thing to note is that one does not require inputs and outputs to be
disjoint subsets of $V$.  This, seemingly innocuous, additional flexibility
is actually quite useful to give parsimonious implementations of
unitaries~\cite{generator04}.  While the restriction to disjoint inputs and
outputs is unnecessary, it has been discussed whether imposing it results
in patterns that are easier to realize physically.  Recent
work~\cite{graphstates,BR04,CMJ04} however, seems to indicate it is not the
case.

\subsection{Pattern combination}
We are interested in how one can combine patterns in order to obtain bigger
ones. 

The first way to combine patterns is by composing them.  
Two patterns $\mfr P_{1}$ and $\mfr P_{2}$ may be composed if
$V_1\cap V_2=O_1=I_2$.  Provided that $\mfr P_1$ has as many outputs
as $\mfr P_2$ has inputs, by renaming the pattern qubits, one can always 
make them composable.
\DE
The composite pattern $\mfr P_{2}\mfr P_{1}$ is defined as:\\ 
--- $V:=V_1\cup V_2$, $I=I_1$, $O=O_2$,\\
--- commands are concatenated.
\ED
The other way of combining patterns is to tensor them.
Two patterns $\mfr P_{1}$ and $\mfr P_{2}$ 
may be tensored if $V_1\cap V_2=\emptyset$.
Again one can always meet this condition by renaming qubits
in a way that these sets are made disjoint.  
\DE
The tensor pattern $\mfr P_{1}\otimes\mfr P_{2}$ is defined as:\\
--- $V=V_1\cup V_2$, $I=I_1\cup I_2$, and $O=O_1\cup O_2$,\\ 
--- commands are concatenated.  
\ED
In contrast to the composition case, all the unions involved here are
disjoint.  Therefore commands from distinct patterns freely commute, since
they apply to disjoint qubits, and when we say that commands have to be
concatenated, this is only for definiteness.
It is routine to verify that the definiteness conditions (D) are preserved
under composition and tensor product.  

Before turning to this matter, we need a clean definition
of what it means for a pattern to implement or to realize a unitary operator,
together with a proof that the way one can combine patterns is reflected in
their interpretations.  This is key to our proof of universality.  

%%%%%%%%%%%%%%%%%%%%%%%%%%%%%%%

\section{The semantics of patterns}
In this section we give a formal operational semantics for the pattern
language as a probabilistic labeled transition system.  We define
deterministic patterns and thereafter concentrate on them.  We show that
deterministic patterns compose.  We give a denotational semantics of
deterministic patterns; from the construction it will be clear that these
two semantics are equivalent.

Besides quantum states, which are non-zero vectors in some Hilbert space
$\hil V$, one needs a classical state recording the outcomes of the
successive measurements one does in a pattern.  If we let $V$ stand for the
finite set of qubits that are still active (i.e.\ not yet measured) and $W$
stands for the set of qubits that have been measured (i.e.\ they are now
just classical bits recording the measurement outcomes), it is natural to
define the computation state space as:
\AR{
\mcl S&:=&\Sigma_{V,W} \hil V\times\ztwo^W.
}
In other words the computation states form a $V,W$-indexed family of
pairs\footnote{These are actually quadruples of the form
$(V,W,q,\Ga)$, unless necessary we will suppress the $V$ and the $W$.} $q$,   
$\Ga$, where $q$ is a quantum state from $\hil V$ and $\Ga$ is a map from
some $W$ to the outcome space $\ztwo$.  We call this classical component
$\Ga$ an \emph{outcome map}, and denote by $\emptyset$ the empty outcome map in
$\ztwo^\emptyset$.  We will treat these states as pairs unless it becomes
important to show how $V$ and $W$ are altered during a computation, as
happens during a measurement.  

\subsection{Operational semantics}
We need some preliminary notation.  For any signal $s$ and classical state 
$\Ga\in\ztwo^W$, such that the domain of $s$ is included in $W$, we take
$s_\Ga$ 
to be the value of $s$ given by the outcome map $\Ga$.  That is to say,
if $s=\sum_I s_i$, then $s_\Ga:=\sum_I\Ga(i)$ where the sum is taken in
$\ztwo$.  
Also if $\Ga\in\ztwo^W$, and $x\in\ztwo$, we define:
\AR{\Ga[x/i](i)=x,\,\Ga[x/i](j)=\Ga(j)\hbox{ for }j\neq i}
which is a map in $\ztwo^{W\cup\ens i}$.

We may now view each of our commands as acting on the state space $\mcl S$,
we have suppressed $V$ and $W$ in the first 4 commands:
\AR{
q,\Ga&\slar{\nq i}&q\otimes\ket{+}_i,\Ga\\
q,\Ga&\slar{\et ij}&\ctR Z_{ij} q,\Ga\\
q,\Ga&\slar{\cx is}&\cx i{s_\Ga} q,\Ga\\
q,\Ga&\slar{\cz is}&\cz i{s_\Ga} q,\Ga\\
V\cup\{i\},W,q,\Ga&\slar{\MS{\al}ist}&V,W\cup\{i\},{\oqbb{\al_\Ga}}_iq,\Ga[0/i]\\
V\cup\{i\},W,q,\Ga&\slar{\MS{\al}ist}&V,W\cup\{i\},{\oqbnb{\al_\Ga}}_iq,\Ga[1/i]
}
where $\al_\Ga=(-1)^{s_\Ga}\al+t_\Ga\pi$ following equation (\ref{msem}).
Note how the measurement moves an index from $V$ to $W$; a qubit once
measured cannot be neasured again.
Suppose $q\in\hil V$, for the above relations to be defined, one needs the
indices $i$, $j$ on which the various command apply to be in $V$.  One also
needs $\Ga$ to contain the domains of $s$ and $t$, so that $s_\Ga$ and
$t_\Ga$ are well-defined.  This will always be the case during the run of a
pattern because of condition (D).

All commands except measurements are deterministic and only modify the
quantum part of the state.  The measurement actions on $\mcl S$ are not
deterministic, so that these are actually binary relations on $\mcl S$, and
modify both the quantum and classical parts of the state.  The usual
convention has it that when one does a measurement the resulting state is
\emph{renormalized} and the probabilities are associated with the
transition.  We do not adhere to this convention here, instead we leave the
states unnormalized.  The reason for this choice of convention is that this
way, the probability of reaching a given state can be read off its norm,
and the overall treatment is simpler.  As we will show later, all the
patterns implementing unitary operators will have the same probability for
all the branches and hence we will not need to carry these probabilities
explicitly.

We introduce an additional command called \emph{signal shifting}:
\AR{
q,\Ga&\slar{\ss is}&q,\Ga[\Ga(i)+s_\Ga/i]
}
It consists in shifting the
measurement outcome at $i$ by the amount $s_\Ga$.  
Note that the $Z$-action leaves measurements globally invariant, in the
sense that $\oqb{\al+\pi},\oqbn{\al+\pi}=\oqbn{\al},\oqb{\al}$.
Thus changing $\al$ to $\al+\pi$ amounts to swapping the outcomes of the
measurements, and one has:
\EQ{\MS\al ist&=&\ss it\,\MS\al is0\label{split}}
and signal shifting allows to dispose of the $Z$ action of a measurement,
resulting sometimes in convenient optimizations of standard forms.

\subsection{Denotational semantics}\label{sub-denot}
Let $\mfr P$ be a pattern with computation space $V$, inputs
$I$, outputs $O$ and command sequence $A_n\ldots A_1$.  
To execute a  
pattern, one starts with some input state $q$ in $\hil I$,
together with the empty outcome map $\emptyset$.  
The input state $q$ is then tensored with as many $\ket{+}$s as there are
non-inputs in $V$ (the $N$ commands), so as to obtain a state in the full
space $\hil V$.  
Then $E$, $M$ and $C$ commands in  $\mfr P$ are applied in sequence from
right to left.  We can
summarize the 
situation as follows: 
\AR{
\xymatrix@=10pt@M=5pt@R=20pt@C=40pt{
{}\hil I\ar[d]\ar@{.>}[rr]
&&
{}\hil O
\\
{}\hil I\times\ztwo^{\emptyset}\ar[r]^{prep}&
{}\hil V\times\ztwo^{\emptyset}\ar[r]^{A_1\ldots A_n\quad}&
{}\hil O\times\ztwo^{V\setminus O}\ar[u]
}
}

If $m$ is the number of measurements, which is also the number of non
outputs, then the run may follow $2^m$ different branches.  Each branch is
associated with a unique binary string $\mathbf{s}$ of length $m$,
representing the classical outcomes of the measurements along that branch,
and a unique \emph{branch map} $A_{\mathbf{s}}$ representing the linear
transformation from $\hil I$ to $\hil O$ along that branch.  This map is
obtained from the operational semantics via the sequence
$(q_i,\Ga_i)$ with $1\leq i\leq n+1$, such that: 
\AR{
q_1,\Ga_1=q\otimes\ket{\hskip-.4ex+\ldots+},\emptyset\\ q_{n+1}=q'\neq0\\
\hbox{and for all }i\leq n:q_i,\Ga_i\slar{A_i}q_{i+1},\Ga_{i+1}.
}

\DE
A pattern $\mfr P$ realizes a map on density matrices $\rho$ given by
$\rho\mapsto \sum_{\mathbf{s}}A_{\mathbf{s}}^{\dag}(\rho) A_{\mathbf{s}}$.
We write $\sem{\mfr P}$ for the map realized by $\mfr P$.
\ED

\PRO
Each pattern realizes a completely positive trace preserving map.
\ORP
\textbf{Proof.} Later on we will show that every pattern can be put in a
semantically 
equivalent form where all the preparations and entanglements appear first,
followed by a sequence of measurements and finally local Pauli corrections.
Hence
branch maps decompose as $A_{\mathbf{s}}=C_{\mathbf{s}}\Pi_{\mathbf{s}}U$,
where $C_{\mathbf{s}}$ is a unitary map over $\hil O$ collecting all
corrections on outputs, $\Pi_{\mathbf{s}}$ is a projection from $\hil V$ to
$\hil O$ representing the particular measurements performed along the
branch, and $U$ is a unitary embedding from $\hil I$ to $\hil V$ collecting
the branch preparations, and entanglements.  Note that $U$ is the same on
all branches.  
Therefore, \\
\AR{
\sum_{\mathbf{s}}A_{\mathbf{s}}^{\dag} A_{\mathbf{s}} & = &
\sum_{\mathbf{s}}U^{\dag}\Pi_{\mathbf{s}}^{\dag}C_{\mathbf{s}}^{\dag}
C_{\mathbf{s}}\Pi_{\mathbf{s}}U\\ & = & 
\sum_{\mathbf{s}}U^{\dag}\Pi_{\mathbf{s}}^{\dag}\Pi_{\mathbf{s}}U\\ & = &
U^{\dag}(\sum_{\mathbf{s}}\Pi_{\mathbf{s}})U\\ & = &
U^{\dag}U = I}
where we have used the fact that $C_{\mathbf{s}}$ is unitary,
$\Pi_{\mathbf{s}}$ is a projection and $U$ is independent of the branches
and is also unitary.  Therefore the map $T(\rho):= \sum_{\mathbf{s}}
A_{\mathbf{s}}(\rho)A_{\mathbf{s}}^{\dag}$ is a trace-preserving
completely-positive map (cptp-map), explicitly given as a Kraus
decomposition.  
\qed
\vskip 0.2cm
Hence the denotational semantics of a pattern is a cptp-map.
In our denotational semantics we view the pattern as defining a map from
the input qubits to the output qubits.  We do not explicitly represent
the result of measuring the final qubits; these may be of interest in some
cases.  Techniques for dealing with classical output explicitly are given
by Selinger~\cite{Selinger04} and Unruh~\cite{Unruh05}.

\DE A pattern is said to be \emph{deterministic} if it realizes a
cptp-map that sends pure states to pure states.  A pattern is said to be
\emph{strongly deterministic} when branch maps are equal.
\ED 

This is equivalent to saying that for a deterministic pattern branch maps
are proportional, that is to say, for all $q\in\hil I$ and all
$\mathbf{s_1}$, $\mathbf{s_2}\in 
\ztwo^n$, $A_{\mathbf{s_1}}(q)$ and $A_{\mathbf{s_2}}(q)$ differ only up to
a scalar.  For a strongly deterministic pattern we have for all
$\mathbf{s_1}$, $\mathbf{s_2}\in \ztwo^n$,
$A_{\mathbf{s_1}}=A_{\mathbf{s_2}}$.  

\PRO
If a pattern is strongly deterministic, then it realizes a unitary
embedding.  
\ORP
\textbf{Proof.} 
Define $T$ to be the map realized by the pattern.  We have
$T=\sum_{\mathbf{s}} A_{\mathbf{s}}^{\dag} A_{\mathbf{s}}$.  Since the
pattern in strongly deterministic all the branch maps are the same.  Define
$A$ to be $2^{n/2}A_{\mathbf{s}}$, then $A$ must be a unitary embedding,
because $A^{\dag} A=I$.  \qed

\subsection{Short examples}
For the rest of paper we assume that all the non-input qubits are prepared
in the state $\ket{+}$ and hence for simplicity we omit the preparation
commands $\nq {I^c}$.  

First we give a quick example of a deterministic pattern that has branches
with different probabilities.  Its type is $V=\ens{1,2}$, $I=O=\ens1$, and
its command sequence is $\M\al2$.  
Therefore, starting with input $q$, one gets two branches:
\AR{
q\otimes\ket+,\emptyset
&\slar{\M\al2}&
\left\{
\begin{array}{l}
\frac12(1+\emi{\al})q,\emptyset[0/2]\\\\
\frac12(1-\emi{\al})q,\emptyset[1/2]
\end{array}
\right.
}
Thus this pattern is indeed deterministic, and implements the identity up
to a global phase, and yet the two branches have respective probabilities
$(1+\cos\al)/2$ and $(1-\cos\al)/2$, which are not equal in general and
hence this pattern is not strongly deterministic.

There is an interesting variation on this first example.  The pattern of
interest, call it $\mfr T$, has the same type as above
with command sequence $\cx{1}{s_2}\M02\et12$.  Again,
$\mfr T$ is deterministic, but not strongly deterministic: the branches have
different probabilities, as in 
the preceding example.  Now, however, these probabilities may depend
on the input.  The associated transformation is a
cptp-map, $T(\rho):=A\rho A^{\dag}+B\rho B^{\dag}$ with: 
\AR{
A:=\MA{1&0\\0&0},\, B:=\MA{0&1\\0&0}
}
One has $A^{\dag} A+B^{\dag} B=I$, so $T$ is indeed a completely positive and
trace-preserving linear map and  
$T(\ket{\psi}\bra{\psi})=\tup{\psi,\psi}\ket{0}\bra{0}$ and clearly for no
unitary $U$ does one have $T(\rho):=U\rho U^{\dag}$.

For our final example, we return to the pattern $\mfr H$, already defined
above.  Consider the pattern with the same qubit space 
$\ens{1,2}$, and the same inputs and outputs $I=\ens1$, $O=\ens2$, as $\mfr H$,
but with a shorter 
command sequence namely $\M 01\et12$.  
Starting with input $q=(a\ket{0}+b\ket{1})\ket+$, one has two computation
branches, branching at $\M 01$:  
\AR{
(a\ket{0}+b\ket{1})\ket+,\emptyset
&\slar{\et12}&
\ost(a\ket{00}+a\ket{01}+b\ket{10}-b\ket{11}),\emptyset\\\\
&\slar{\M 01}& 
\left\{
\begin{array}{l}
\frac12((a+b)\ket{0}+(a-b)\ket{1}),\emptyset[0/1]\\\\
\frac12((a-b)\ket{0}+(a+b)\ket{1}),\emptyset[1/1]
\end{array}
\right.
}
and since $\norm{a+b}^2+\norm{a-b}^2=2(\norm a^2+\norm b^2)$,
both transitions happen with equal probabilities $\frac12$.  
Both branches end up with non proportional outputs, so the pattern is \emph{not}
deterministic.  However, if one applies the local correction $\Cx2$ on
either of the branches' ends, both outputs will be made to coincide.  If we
choose to let the correction apply to the second branch, we obtain the pattern
$\mfr H$, already defined.  We have just proved $H=U_{\mfr H}$,
that is to say $\mfr H$ realizes the Hadamard operator.

\subsection{Compositionality of the Denotational Semantics}
With our definitions in place, we will show that the denotational semantics
is compositional.

\TH
For two patterns $\mfr{P}_1$ and $\mfr{P}_2$ we have
$\sem{\mfr{P}_1\mfr{P}_2} =
\sem{\mfr{P}_2}\sem{\mfr{P}_1}$ and 
$\sem{\mfr{P}_1\otimes\mfr{P}_2} =
\sem{\mfr{P}_2}\otimes\sem{\mfr{P}_1}.$
\HT
\textbf{Proof.}
Recall that two patterns $\mfr P_1$, $\mfr P_2$ may be combined by
composition provided $\mfr P_1$ has as many outputs as $\mfr P_2$ has
inputs.  Suppose 
this is the case, and suppose further that $\mfr P_1$ and $\mfr P_2$
respectively realize some cptp-maps $T_1$ and $T_2$.  We need to show that
the composite pattern $\mfr P_2\mfr P_1$ realizes $T_2T_1$.

Indeed, the two diagrams representing branches in $\mfr P_1$ and $\mfr P_2$:
{\footnotesize
\AR{
\xymatrix@=10pt@M=3pt@R=20pt@C=7pt{
{}\hil {I_1}\ar[d]\ar@{.>}[rr]
&&
{}\hil {O_1}\ar@{=}
&
{}\hil {I_2}\ar[d]\ar@{.>}[rr]
&&
{}\hil {O_2}
\\
{}\hil {I_1}\times\ztwo^{\emptyset}\ar[r]^{p_1}&
{}\hil {V_1}\times\ztwo^{\emptyset}\ar[r]^{}&
{}\hil {O_1}\times\ztwo^{V_1\setminus O_1}\ar[u]
&
{}\hil {I_2}\times\ztwo^{\emptyset}\ar[r]^{p_2}&
{}\hil {V_2}\times\ztwo^{\emptyset}\ar[r]^{}&
{}\hil {O_2}\times\ztwo^{V_2\setminus O_2}\ar[u]
}
}
}\\
can be pasted together, since $O_1=I_2$, and $\hil {O_1}=\hil {I_2}$.  
But then, it is enough to notice 1) that preparation steps $p_2$ in $\mfr P_2$
commute with all actions in $\mfr P_1$ since they apply on disjoint sets
of qubits, and 2) that no action taken in $\mfr P_2$ depends on 
the measurements outcomes in $\mfr P_1$.  It follows that 
the pasted diagram describes the same branches as does
the one associated to the composite $\mfr P_2\mfr P_1$.  

A similar argument applies to the case of a tensor combination,
and one has that $\mfr P_2\otimes\mfr P_1$ realizes $T_2\otimes T_1$.
\qed
\vskip 0.2cm

If one wanted to give a categorical treatment\footnote{The rest of the
  paragraph can be omitted without loss of continuity.} one can define a
category where the objects are finite sets representing the input and
output qubits and the morphisms are the patterns.  This is clearly a
monoidal category with our tensor operation as the monoidal structure.  One
can show that the denotational semantics gives a monoidal functor into the
category of superoperators or into any suitably enriched strongly compact
closed category~\cite{AbrCoe04} or dagger category~\cite{Selinger05}.  It
would be very interesting to explore exactly what additional categorical
structures are required to interpret the measurement calculus presented
below.  Duncan Ross\cite{Duncan05} has skectched a polycategorical
presentation of our measurement calculus.

%%%%%%%%%%%%%%%%%%%%%%%%%%%%%%%

\section{Universality}\label{s:univ}
Define the two following patterns on $V=\ens{1,2}$: 
\EQ{
\mfr \G(\al)&:=&\cx 2{s_1}\M{{-\al}}1\et 12\\
\ctR{\mfr Z}&:=&\et 12 
}
with $I=\ens1$, $O=\ens2$ in the first pattern, and
$I=O=\ens{1,2}$ in the second.  Note that the second pattern
does have overlapping inputs and outputs.  

\PRO
The patterns $\mfr \G(\al)$ and $\ctR{\mfr Z}$ are universal.
\ORP
\textbf{Proof.}
First, we claim $\mfr \G(\al)$ and $\ctR{\mfr Z}$ respectively
realize $\G(\al)$ and $\ctR{Z}$, with:
\AR{ 
\G(\al)&:=&\ost\MA{1&\ei\al\\1&-\ei\al}
}
We have already seen in our example that $\mfr\G(0)=\mfr H$ implements
$H=\G(0)$, thus we already know this in the particular 
case where $\al=0$.  The general case follows by the same kind of computation.%
\footnote{Equivalently, this follows from $\G(\al)=HP(\al)$, with $P(\al)=\MA{1&0\\0&\ei\al}$ and: 
$$
\cx 2{s_1}\M{{-\al}}1\et 12=
\cx 2{s_1}\M{{0}}1P(\al)_1\et 12=\mfr H P(\al)_1.
$$
}
The case of $\ctR Z$ is obvious.\\
Second, we know that these unitaries form a universal set for
$\otimes^n\ctwo$~\cite{generator04}.  Therefore, from the preceding
section, we infer that combining the corresponding patterns
will generate patterns realizing any unitary in $\otimes^n\ctwo$.
\qed
\vskip 0.2cm

These patterns are indeed among the simplest possible.  As a consequence, in
the section devoted to examples, we will find that our implementations
often have lower space complexity than the traditional implementations.

Remarkably, in our set of generators, one finds a single measurement and a
single dependency, which occurs in the correction phase of $\mfr \G(\al)$.
Clearly one needs at least one measurement, since patterns without
measurements can only implement unitaries in the Clifford group.  It is also
true that dependencies are needed for universality, but we have to wait for
the development of the measurement calculus in the next section to give a
proof of this fact.

%%%%%%%%%%%%%%%%%%%%%%%%%%%%%%%

\section{The measurement calculus}\label{s-cal}
We turn to the next important matter of the paper, namely standardization.
The idea is quite simple.  It is enough to provide local pattern-rewrite
rules pushing $E$s to the beginning of the pattern and $C$s to the end.
The crucial point is to justify using the equations as rewrite rules.  

\subsection{The equations}

The expressions appearing as commands are all linear operators on Hilbert
space.  At first glance, the appropriate equality between commands is
equality as operators.  For the deterministic commands, the equality that
we consider is indeed equality as operators.  This equality implies
equality in the denotational semantics.  However, for measurement commands
one needs a stricter definition for equality in order to be able to apply
them as rewriting rules.  Essentially we have to take into the account the
effect of different branches that might result from the measurement
process.  The precise definition is below.

\DE
Consider two patterns $\mfr P$ and $\mfr P'$ we define $\mfr P= \mfr P'$ if
and only if for any branch $s$, we have $A^{\mfr P}_s=A^{\mfr P'}_s$, where
$A^{\mfr P}_s$ and $A^{\mfr P'}_s$ are the branch map $A_s$ defined in
Section \ref{sub-denot}. 
\ED

The first set of equations gives the means to propagate local Pauli
corrections through the entangling operator $\et ij$.  
\EQ{
\et ij \cx is&=&\cx is\cz js\et ij\label{ecx}\\
\et ij \cx js&=&\cx js\cz is\et ij\label{ecx2}\\
\et ij \cz is&=&\cz is\et ij\label{ecz}\\
\et ij \cz js&=&\cz js\et ij\label{ecz2}
}
These equations are easy to verify and are natural since $\et ij$ belongs
to the Clifford group, and therefore maps under conjugation the Pauli group
to itself.  Note that, despite the symmetry of the $E_{ij}$ operator
\emph{qua} operator, we have to consider all the cases, since the rewrite
system defined below does not allow one to rewrite $E_{ij}$ to $E_{ji}$.
If we did allow this the reqrite process could loop forever.

A second set of equations allows one to push corrections through
measurements acting on the same qubit.  Again there are two cases: 
\EQ{ 
\MS\al ist\cx ir&=&\MS\al i{s+r}{t}\label{mcx}\\
\MS\al ist\cz ir&=&\MS\al i{s}{t+r}\label{mcz}
}
These equations follow easily from equations (\ref{e:simp1}) and
(\ref{e:simp2}).  
They express the fact that the 
measurements $\Ms\al i$ are closed under conjugation by the Pauli group, 
very much like equations~(\ref{ecx}),(\ref{ecx2}),(\ref{ecz})
and~(\ref{ecz2}) express the fact that the Pauli group is closed under 
conjugation by the entanglements $\et ij$.

Define the following convenient abbreviations:
\AR{
\ms\al is:=\MS\al is0,\,
\MS\al i{}t:=\MS\al i0t,\,
\Ms\al i:=\MS\al i00,\\
\Ms xi:=\Ms0i,\,
\Ms yi:=\Ms\pit i
}
Particular cases of the equations above are:
\AR{ 
\M xi\cx is&=&\M xi\\
\M yi\cx is&=&\ms yis &=&\MS yi{}s &=&\M yi\cz is
}
The first equation, follows from the fact that ${-0}=0$, so the $X$ action
on $\M xi$ 
is trivial; the second equation, is because 
${-\pit}$ is equal $\pit+\pi$ modulo $2\pi$, 
and therefore the $X$ and $Z$ actions coincide on $\M yi$.
So we obtain the following:
\EQ{ 
\MS xist&=&\MS xi{}t\label{mx}\\
\MS yist&=&\MS yi{}{s+t}\label{my}
}
which we will use later to prove that patterns with measurements
of the form $M^x$ and $M^y$ may only realize unitaries in the 
Clifford group.

\subsection{The rewrite rules}
We now define a set of rewrite rules, obtained by orienting the equations
above\footnote{Recall that patterns are executed from right to left.}:  
\AR{
\et ij\cx is&\Rar&\cx is\cz js\et ij&\quad\hbox{}EX\\
\et ij\cx js&\Rar&\cx js\cz is\et ij&\quad\hbox{}EX\\
\et ij\cz is&\Rar&\cz is\et ij&\quad\hbox{}EZ\\
\et ij\cz js&\Rar&\cz js\et ij&\quad\hbox{}EZ\\
\MS\al is{t}\cx i{r}&\Rar&\MS\al i{s+r}{t}&\quad\hbox{}MX\\
\MS\al is{t}\cz i{r}&\Rar&\MS\al i{s}{r+t}&\quad\hbox{}MZ }
to which we need to add the \emph{free commutation rules}, 
obtained when commands operate on disjoint sets of qubits: 
\AR{
\et ij\CO{\vec k}&\Rar&\CO{\vec k}\et ij&\quad\hbox{where $A$ is not an
  entanglement}\\
\CO{\vec k}\cx is&\Rar&\cx is\CO{\vec k}&\quad\hbox{where $A$ is not 
a correction}\\
\CO{\vec k}\cz is&\Rar&\cz is\CO{\vec k}&\quad\hbox{where $A$ is not a
  correction}
} 
where $\vec k$ represent the qubits acted upon by command $A$,
and are supposed to be distinct from $i$ and $j$.  Clearly these rules
could be reversed since they hold as equations but we are orienting them
this way in order to obtain termination.

Condition (D) is easily seen to be preserved under rewriting.

Under rewriting, the computation space, inputs and outputs remain the same,
and so do the entanglement commands.  Measurements might be modified, but
there is still the same number of them, and they still act on the same
qubits.  The only induced modifications concern local corrections and
dependencies.  If there was no dependency at the start, none will be
created in the rewriting process.  

In order to obtain rewrite rules, it was essential that the entangling
command ($\ctR Z$) belongs to the normalizer of the Pauli group.  The point
is that the Pauli operators are the correction operators and they can be
dependent, thus we can commute the entangling commands to the beginning
without inheriting any dependency.  Therefore the entanglement resource can
indeed be prepared at the outset of the computation.

\subsection{Standardization}

Write $\mfr P\Rar\mfr P'$, respectively $\mfr P\Rar\st\mfr P'$, if both
patterns have the same type, and one obtains the command sequence of $\mfr P'$
from the command sequence of $\mfr P$ by applying one, respectively any
number, of the rewrite rules of the previous section.  We say that $\mfr P$ is
\emph{standard} if for no $\mfr P'$, $\mfr P\Rar\mfr P'$ and the procedure
of writing a pattern to standard form is called standardization\footnote{We
  use the word ``standardization'' instead of the more usual
  ``normalization'' in order not to cause terminological confusion with the
  physicists' notion of normalization.}.

One of the most important results about the rewrite system is that it has
the desirable properties of determinacy (confluence) and termination
(standardization).  In other words, we will show that for all $\mfr P$,
there exists a unique standard $\mfr P'$, such that $\mfr P\Rar\st\mfr P'$.
It is, of course, crucial that the standardization process leaves the
semantics of patterns invariant.  This is the subject of the next simple,
but important, proposition,

\PRO  Whenever $\mfr P\Rar\st\mfr P'$, $\sem{\mfr P}=\sem{\mfr P'}$.  \ORP
\textbf{Proof.}
It is enough to prove it when $\mfr P\Rar\mfr P'$.  The
first group of rewrites has been proved to be sound in the preceding
subsections, while the free commutation rules are obviously sound.  \qed 
\vskip 0.2cm

We now begin the main proof of this section.  First, we prove termination.
\TH[Termination] \label{t-term} All rewriting sequences beginning with a
pattern $\mfr P$ terminate after finitely many steps.  For our rewrite
system, this implies that for
all $\mfr P$ there exist finitely many $\mfr P'$ such that $\mfr
P\Rar\st\mfr P'$ where the $\mfr P'$ are standard.  \HT

\textbf{Proof.}  
Suppose $\mfr P$ has command sequence $A_n\ldots A_1$; so the number of
commands is $n$.  Let $e\leq n$ be the number of $E$ commands in $\mfr P$.
As we have noted earlier, this number is invariant under $\Rar$.  Moreover
$E$ commands in $\mfr P$ can be ordered by increasing depth, read from
right to left, and this order, written $<_E$, is also invariant, since $EE$
commutations are forbidden explicitly in the free commutation rules.

Define the following depth function $d$ on $E$ and $C$ commands in $\mfr P$:
\AR{
d(A_i)=
\left\{\begin{array}{ll}
i&\hbox{if }A_i=\et jk\\
n-i&\hbox{if }A_i=C_j
\end{array}
\right.
}
Define further the following sequence of length $e$, $d_E(\mfr
P)(i)$ is the depth of the $E$-command of rank $i$ according to $<_E$.  By
construction this sequence is strictly increasing.  Finally, we define the
measure $m(\mfr P):=(d_E(\mfr P),d_C({\mfr P}))$ with: \AR{ d_C({\mfr
    P})&=&\sum_{C\in\mfr P}d(C) } We claim the measure we just defined
decreases lexicographically under rewriting, in other words $\mfr P\Rar\mfr
P'$ implies $m(\mfr P)>m(\mfr P')$, where $<$ is the lexicographic ordering
on $\mbb N^{e+1}$.

To clarify these definitions, consider the following example.  Suppose
$\mfr P$'s command sequence is of the form $EXZE$, then $e=2$, $d_E(\mfr
P)=(1,4)$, and $m(\mfr P)=(1,4,3)$.  For the command sequence $EEX$ we get
that $e=2$, $d_E(\mfr P)=(2,3)$ and $m(\mfr P)=(2,3,2)$.  Now, if one
considers the rewrite $EEX\Rar EXZE$, the measure of the left hand side is
$(2,3,2)$, while the measure of the right hand side, as said, is $(1,4,3)$,
and indeed $(2,3,2)>(1,4,3)$.  Intuitively the reason is clear: the $C$s
are being pushed to the left, thus decreasing the depths of $E$s, and
concomitantly, the value of $d_E$.
 
Let us now consider all cases starting with an $EC$ rewrite.  Suppose the
$E$ command under rewrite has depth $d$ and rank $i$ in the order $<_E$.
Then all $E$s of smaller rank have same depth in the right hand side, while
$E$ has now depth $d-1$ and still rank $i$.  So the right hand side has a
strictly smaller measure.  Note that when $C=X$, because of the creation of
a $Z$ (see the example above), the last element of $m(\mfr P)$ may
increase, and for the same reason all elements of index $j>i$ in $d_E(\mfr
P)$ may increase.  This is why we are working with a lexicographical
ordering.

Suppose now one does an $MC$ rewrite, then $d_C(\mfr P)$ strictly
decreases, since one correction is absorbed, while all $E$ commands have
equal or smaller depths.  Again the measure strictly decreases.

Next, suppose one does an $EA$ rewrite, and the $E$ command under rewrite
has depth $d$ and rank $i$.  Then it has depth $d-1$ in the right hand side,
and all other $E$ commands have invariant depths, since we forbade the case
when $A$ is itself an $E$.  It follows that the measure strictly decreases.

Finally, upon an $AC$ rewrite, all $E$ commands have invariant depth,
except possibly one which has smaller depth in the case $A=E$, and
$d_C(\mfr P)$ decreases strictly because we forbade the case where $A=C$.
Again the claim follows.

So all rewrites decrease our ordinal measure, and therefore all sequences
of rewrites are finite, and since the system is finitely branching (there
are no more than $n$ possible single step rewrites on a given sequence of
length $n$), we get the statement of the theorem.  

The final statement of the theorem follows from the fact that we have
finitely many rules so the system is \emph{finitely branching}.  In any
finitely branching rewrite system with the property that every rewrite
sequence terminates, it is clearly true that there can be only finitely many
standard forms.
\qed \vskip 0.2cm

The next theorem establishes the important determinacy property and
furthermore shows that the standard patterns have a certain canonical
form which we call the NEMC form.  The precise definition is:
\DE
A pattern has a \emph{NEMC} form if its commands occur in the order of $N$s
first, then $E$s , then $M$s, and finally $C$s.   
\ED
We will usually just say ``EMC'' form since we can assume that all the
auxiliary qubits are prepared in the $\ket{+}$ state we usually just elide
these $N$ commands.

\TH[Confluence] \label{t-conf}
For all $\mfr P$, there exists a unique standard $\mfr P'$, such that $\mfr
P\Rar\st\mfr P'$, and $\mfr P'$ is in EMC form.  
\HT
\textbf{Proof.} Since the rewriting system is terminating, confluence
follows from local 
confluence~\footnote{This means that whenever two rewrite rules can be
  applied to a term $t$ yielding $t_1$ and $t_2$, one can rewrite both
  $t_1$ and $t_2$ to a common third term $t_3$, possibly in many steps.} by
Newman's lemma, see, for example,~\cite{Barendregt84}.  The uniqueness of
the standard is form an immediate consequence.

We look for critical pairs, that is occurrences of three successive
commands where two rules can be applied simultaneously.  One finds that
there are only five types of critical pairs, of these the three involve the
$N$ command, these are of the form:  $NMC$, $NEC$ and $NEM$; and the
remaining two are: $\et ijM_kC_k$ with $i$, $j$ and $k$ all distinct,  
$\et ijM_kC_{l}$ with $k$ and $l$ distinct.
In all cases local confluence is easily verified.

Suppose now ${\mfr P'}$ does not satisfy the EMC form conditions.  Then,
either there is a pattern $EA$ with $A$ not of type $E$, or there is a
pattern $AC$ with $A$ not of type $C$.  In the former case, $E$ and $A$
must operate on overlapping qubits, else one may apply a free commutation
rule, and $A$ may not be a $C$ since in this case one may apply an $EC$
rewrite.  The only remaining case is when $A$ is of type $M$, overlapping
$E$'s qubits, but this is what condition (D1) forbids, and since (D1) is
preserved under rewriting, this contradicts the assumption.  The latter
case is even simpler.  \qed

We have shown that under rewriting any pattern can be put in EMC form,
which is what we wanted.  We actually proved more, namely that the standard
form obtained is unique.  However, one has to be a bit careful about the
significance of this additional piece of information.  Note first that
uniqueness is obtained because we dropped the $CC$ and $EE$ free
commutations, thus having a rigid notion of command sequence.  One cannot
put them back as rewrite rules, since they obviously ruin termination and
uniqueness of standard forms.

A reasonable thing to do, would be to take this set of equations as
generating an equivalence relation on command sequences, call it $\equiv$,
and hope to strengthen the results obtained so far, by proving that all
reachable standard forms are equivalent.

But this is too naive a strategy, since 
$\et12\Cx1\Cx2\equiv\et12\Cx2\Cx1$, and: 
\AR{ \et12\cx1s\cx2t
&\Rar\st&\cx1s\cz2s\cx2t\cz1t\et12\\
&\equiv&\cx1s\cz1t\cz2s\cx2t\et12 
} 
obtaining an expression which is not symmetric in $1$ and $2$.  To conclude,
one has to extend $\equiv$ to include the additional equivalence
$\cx1s\cz1t\equiv\cz1t\cx1s$, which fortunately is sound since these two
operators are equal up to a global phase.  Thus, these are all equivalent in
our semantics of patterns.  We summarize this discussion as follows.
\DE
We define an equivalence relation $\equiv$ on patterns by taking all the
rewrite rules as equations and adding the equation
$\cx1s\cz1t\equiv\cz1t\cx1s$ and generating the smallest equivalence
relation. 
\ED
With this definition we can state the following proposition.
\PRO
All patterns that are equivalent by $\equiv$ are equal in the denotational semantics. 
\ORP
This $\equiv$ relation preserves both the type (the $(V,I,O)$ triple) and the underlying entanglement graph. So clearly semantic equality does not entail equality up to $\equiv$. In fact, by composing teleportation patterns one obtains infinitely many patterns for the identity which are all different up to $\equiv$. One may wonder whether two patterns with same semantics, type and underlying entanglement graph are necessarily equal up to $\equiv$. This is not true either. One has $J(\al)J(0)J(\ba)=J(\al+\ba)=J(\ba)J(0)J(\al)$ (where $J(\al)$ is defined in Section~\ref{s:univ}), and this readily gives a counter-example.

We can now formally describe a simple standardization algorithm.
\AL \label{A-graph}
\emph{Input}: A pattern $\mfr P$ on $|V|=N$ qubits with command sequence
$A_M \cdots A_1$.  
\\ \emph{Output}: An equivalent pattern $\mfr P'$ in NEMC form. 
\begin{enumerate}
\item Commute all the preparation commands (new qubits) to the right side.
\item Commute all the correction commands to the left side using the EC and
  MC rewriting rules.   
\item Commute all the entanglement commands to the right side after the
  preparation commands. 
\end{enumerate}
\LA

Note that since each qubit can be entangled with at most $N-1$ other
qubits, and can be measured or corrected only once, we have $O(N^2)$
entanglement commands and $O(N)$ measurement commands.  According to the
definiteness condition, no command acts on a qubit not yet prepared, hence
the first step of the above algorithm is based on trivial commuting rules;
the same is true for the last step as no entanglement command can act on a
qubit that has been measured.  Both steps can be done in $O(N^2)$.  The
real complexity of the algorithm comes from the second step and the $EX$
commuting rule.  In the worst case scenario, commuting an $X$ correction to
the left might create $O(N^2)$ other $Z$ corrections, each of which has to
be commuted to the left themselves.  Thus one can have at most $O(N^3)$ new
corrections, each of which has to be commuted past $O(N^2)$ measurement or
entanglement commands.  Therefore the second step, and hence the algorithm,
has a worst case complexity of $O(N^5)$.

We conclude this subsection by emphasizing the importance of the EMC form.  
Since the entanglement can always be done first, we can always derive the
entanglement resource needed for the whole computation right at the
beginning.  After that only local operations will be performed.  This will
separate the analysis of entanglement resource requirements from the
classical control.  Furthermore, this makes it possible to extract the
maximal parallelism for the execution of the pattern since the necessary
dependecies are explicitly expressed, see the example in
section~\ref{s:examples} for further discussion.  Finally, the EMC form
provides us with tools to prove general theorems about patterns, such as
the fact that they always compute cptp-maps and the expressiveness theorems
of section~\ref{s:expr}.

\subsection{Signal shifting}
One can extend the calculus to include the signal shifting command $\ss it$.
This allows one to dispose of dependencies induced by the $Z$-action, and
obtain sometimes standard patterns with smaller computational depth
complexity, as we will see in the next section which is devoted to examples.  
\AR{
\MS\al ist&\Rar&\ss it\ms\al is\\
\cx js\ss it&\Rar& \ss it \cx j{s[t+s_i/s_i]}\\
\cz js\ss it&\Rar& \ss it \cz j{s[t+s_i/s_i]}\\
\MS\al jst\ss ir&\Rar&\ss ir\, \MS\al j{s[r+s_i/s_i]}{t[r+s_i/s_i]}\\
\ss is \ss jt&\Rar&\ss jt \ss i{s[t+s_j/s_j]}
}
where $s[t/s_i]$ denotes the substitution of $s_i$ with
$t$ in $s$, $s$, $t$ being signals.  Note that when we write a $t$
explicitly on the upper left of an $M$, we mean that $t\not=0$.
The first additional rewrite rule was already introduced
as equation (\ref{split}), while the other ones merely propagate
the signal shift.  Clearly one can dispose of $\ss it$ when it hits
the end of the pattern command sequence. We will refer to this new set of rules as $\Rar_S$. Note that we always apply first the standardization rules and then signal shifting, hence we do not need any commutation rule for $E$ and $S$ commands. 

It is important to note that both theorem \ref{t-term} and \ref{t-conf}
still hold for this extended rewriting system.  In order to prove
termination one can start with the EMC form and then adapt the proof of
Theorem \ref{t-term} by defining a depth function for a signal shift
similar to the depth of a correction command.  As with the correction,
signal shifts can also be commuted to the left hand side of a command
sequence.  Now our measure can be modified to account for the new signal
shifting terms and shown to be decreasing under each step of signal
shifting.  Confluence can be also proved from local confluence using again
Newman's Lemma~\cite{Barendregt84}.  One typical critical pair is $\mL \al
j t S_i^s$ where $i$ appears in the domain of signal $t$ and hence the
signal shifting command $S_i^s$ will have an effect on the measurement.
Now there are two possible ways to rewrite this pair, first, commute the
signal shifting command and then replace the left signal of the measurement
with its own signal shifting command:
\AR{
  \mL \al j t \, S_i^s \, &\Rar& \, S_i^s\, \mL \al j {t+s} \\
  &\Rar&\, S_i^s\, S_j^{s+t}\, \m \al j } The other way is to first replace
the left signal of the measurement and then commute the signal shifting
command: \AR{
  \mL \al j t \, S_i^s \, &\Rar&\,  S_j^t \,\m \al j  \, S_i^s\\
  &\Rar&\, S_j^t\, S_i^{s}\, \m \al j } Now one more step of rewriting on
the last equation will give us the same result for both choices.  \AR{
  S_j^t\, S_i^{s}\, \m \al j \,\Rar \, S_i^s\, S_j^{s+t}\, \m \al j } All
other critical terms can be dealt with similarly.

%%%%%%%%%%%%%%%%%%%%%%%%%%%%%%%

\section{Examples}\label{s:examples}
In this section we develop some examples illustrating pattern
composition, pattern standardization, and signal shifting.  We compare our
implementations with the implementations given in the reference
paper~\cite{mqqcs}.  To combine patterns 
one needs to rename their qubits as we already noted.  We use the 
following concrete notation: if $\mfr P$ is a pattern over
$\ens{1,\ldots,n}$,  
and $f$ is an injection, we write $\mfr P(f(1),\ldots,f(n))$
for the same pattern with qubits renamed according to $f$.  We also
write $\mfr P_2\circ\mfr P_1$ for pattern composition, in order to make it
more readable. 
Finally we define the \emph{computational depth complexity} to be the
number of measurement rounds plus one final correction round.  More details
on depth complexity, specially on the preparation depth, \ie\,depth of the
entanglement commands, can be found in \cite{BK06}. 

\subsubsection*{Teleportation.}
Consider the composite pattern $\mfr\G(\ba)(2,3)\circ\mfr\G(\al)(1,2)$ with
computation space $\ens{1,2,3}$, inputs $\ens{1}$, and outputs $\ens{3}$.  
We run our standardization procedure so as to obtain an equivalent standard
pattern:  
\AR{
\mfr \G(\ba)(2,3)\circ\mfr \G(\al)(1,2)&=&
\cx3{s_{2}}\Ms{-\ba}2\mybox{\et23\cx2{s_{1}}}\Ms{-\al}1\et12
\\&\Rar_{EX}&
\cx3{s_{2}}\mybox{\Ms{-\ba}2\cx2{s_1}}\cz3{s_1}\Ms{-\al}1\et23\et12
\\&\Rar_{MX}&
\cx3{s_{2}}\cz3{s_{1}}\ms{-\ba}2{s_{1}}\Ms{-\al}1\et23\et12
}
Let us call the pattern just obtained $\mfr \G(\al,\ba)$.  If we take as
a special case $\al=\ba=0$, we get: 
\AR{
\cx3{s_2}\cz3{s_1}\Ms x2\Ms x1\et23\et12
}
and since we know that $\mfr \G(0)$ implements $H$ and $H^2=I$, we conclude
that this pattern 
implements the identity, or in other words it teleports qubit $1$ to
qubit $3$.  As it happens, this pattern obtained by self-composition, is the
same as the one given in the reference paper~\cite[p.14]{mqqcs}.  

\subsubsection*{$x$-rotation.}
Here is the reference implementation of an $x$-rotation~\cite[p.17]{mqqcs},
$R_x(\al)$: 
\AR{
\cx3{s_2}\cz3{s_1}\ms{-\al}2{s_1}\Ms x1\et23\et12
}
with type $\ens{1,2,3}$, $\ens{1}$, and $\ens{3}$.  There is a natural
question which one might call the recognition problem, namely how does one
know this is implementing $R_x(\al)$~? Of course there is the brute force
answer to that, which we applied to compute our simpler patterns, and which
consists in computing down all the four possible branches generated by the
measurements at qubits $1$ and $2$.  Another possibility is to use the
stabilizer formalism as explained in the reference paper~\cite{mqqcs}.  Yet
another possibility is to use \emph{pattern composition}, as we did before,
and this is what we are going to do.

We know that $R_x(\al)=\G({\al})H$ up to a global phase, hence the
composite pattern $\mfr \G({\al})(2,3)\circ\mfr H(1,2)$ implements
$R_{x}(\al)$.  
Now we may standardize it:
\AR{
\mfr \G({\al})(2,3)\circ\mfr H(1,2)&=&
\cx3{s_2}\Ms{-\al}2\mybox{\et23\cx2{s_1}}\Ms x1\et12\\
&\Rar_{EX}&
\cx3{s_2}\cz3{s_1}\mybox{\Ms{-\al}2\cx2{s_1}}\Ms x1\et23\et12\\
&\Rar_{MX}&
\cx3{s_2}\cz3{s_1}\ms{-\al}2{s_1}\Ms x1\et23\et12\\
}
obtaining exactly the implementation above.
Since our calculus preserves the semantics, we deduce that
the implementation is correct.

\subsubsection*{$z$-rotation.}
Now, we have a method here for synthesizing further implementations.
Let us replay it with another rotation $R_z(\al)$.  
Again we know that $R_z(\al)=HR_x(\al)H$, and we already know how to
implement both components $H$ and $R_x(\al)$.  

So we start with the pattern
$\mfr H(4,5)\circ\mfr R_x(\al)(2,3,4)\circ\mfr H(1,2)$ and standardize it:
\AR{
\mfr H(4,5)\circ\mfr R_x(\al)(2,3,4)\circ\mfr H(1,2)=\\
\mfr H(4,5)
\cx4{s_3}\cz4{s_2}
\MS\al3{1+s_2}{}
\Ms x2
\et34
\mybox{\et23
\cx2{s_1}}
\Ms x1
\et12
&\Rar_{EX}&\\
\mfr H(4,5)
\cx4{s_3}\cz4{s_2}\MS\al3{1+s_2}{}\Ms x2
\cx2{s_1}
\mybox{\et34\cz3{s_1}}
\Ms x1
\et1{23}
&\Rar_{EZ}&\\
\mfr H(4,5)
\cx4{s_3}\cz4{s_2}\MS\al3{1+s_2}{}\cz3{s_1}
\mybox{\Ms x2\cx2{s_1}}
\Ms x1
\et1{234}
&\Rar_{MX}&\\
\mfr H(4,5)
\cx4{s_3}\cz4{s_2}
\mybox{\MS\al3{1+s_2}{}\cz3{s_1}}
\Ms x2
\Ms x1
\et1{234}
&\Rar_{MZ}&\\
\cx5{s_4}
\Ms x4
\mybox{\et45\cx4{s_3}}\cz4{s_2}
\MS\al3{1+s_2}{s_1}
\Ms x2
\Ms x1
\et1{234}
&\Rar_{EX}&\\
\cx5{s_4}
\cz5{s_3}
\mybox{\Ms x4\cx4{s_3}}
\cz4{s_2}
\MS\al3{1+s_2}{s_1}
\Ms x2
\Ms x1
\et1{2345}
&\Rar_{MX}&\\
\cx5{s_4}
\cz5{s_3}
\mybox{\MS x4{s_3}{}\cz4{s_2}}
\MS\al3{1+s_2}{s_1}
\Ms x2
\Ms x1
\et1{2345}
&\Rar_{MZ}&\\
\cx5{s_4}
\cz5{s_3}
\MS x4{s_3}{s_2}
\MS\al3{1+s_2}{s_1}
\Ms x2
\Ms x1
\et1{2345}
}
To aid reading $\et23\et12$ is shortened to $\et1{23}$, $\et12\et23\et34$
to $\et1{234}$, and $\MS{\al}i{1+s}{t}$ is used as shorthand for
$\MS{-\al}i{s}{t}$.  

Here for the first time, we see $MZ$ rewritings, inducing the $Z$-action on
measurements.  The resulting standardized pattern can therefore be rewritten
further using the extended calculus: 
\AR{
\cx5{s_4}
\cz5{s_3}
\MS x4{s_3}{s_2}
\MS\al3{1+s_2}{s_1}
\Ms x2
\Ms x1
\et1{2345}
&\Rar_{S}&\\
\cx5{s_2+s_4}\cz5{s_1+s_3}
\Ms x4
\MS\al3{1+s_2}{}
\Ms x2\Ms x1
\et1{2345}
}
obtaining the pattern given in the reference paper~\cite[p.5]{mqqcs}.

However, just as in the case of the $R_x$ rotation, we also
have $R_z(\al)=H\G({\al})$ up to a global phase, 
hence the pattern $\mfr H(2,3)\mfr \G({\al})(1,2)$ also 
implements $R_{z}(\al)$, and we may standardize it:
\AR{
\mfr H(2,3)\circ\mfr \G({\al})(1,2)&=&
\cx3{s_2}\Ms x2
\mybox{\et23\cx2{s_1}}
\Ms{-\al}1\et12
\\&\Rar_{EX}&
\cx3{s_2}
\cz3{s_1}
\mybox{\Ms x2
\cx2{s_1}
}\Ms{-\al}1\et1{23}
\\&\Rar_{MX}&
\cx3{s_2}
\cz3{s_1}
\Ms x2{}{}
\Ms{-\al}1\et1{23}
}
obtaining a 3 qubit standard pattern for the $z$-rotation, which is simpler
than the preceding one, because it is based on the $\mfr \G(\al)$ generators.  
Since the $z$-rotation $R_z(\al)$ is the same as the phase operator:
\AR{P(\al)=\MA{1&0\\0&\ei\al}}
up to a global phase, we also obtain with the same pattern an implementation
of the phase operator.  In particular, if $\al=\pit$, using
the extended calculus, we get the following pattern for $P(\pit)$:
$\cx3{s_2}\cz3{s_1+1}\Ms x2\Ms y1\et1{23}$.

\subsubsection*{General rotation.}
The realization of a general rotation based
on the Euler decomposition of rotations as $R_x(\ga)R_z(\ba)R_x(\al)$,
would results in a 7 qubit pattern.  We get a 5 qubit implementation 
based on the $\G(\al)$ decomposition~\cite{generator04}:
\AR{
R(\al,\ba,\ga)&=&\G(0) \G(-\al)\G(-\ba)\G(-\ga)
}
(The parameter angles are inverted to make the computation
below more readable.) The extended standardization procedure yields:
\AR{
\mfr \G(0)(4,5)
\mfr \G(-\al)(3,4)
\mfr \G(-\ba)(2,3)
\mfr \G(-\ga)(1,2)
&=&\\
\cx5{s_4}\Ms{0}4\et45
\cx4{s_3}\Ms{\al}3\et34
\cx3{s_2}\Ms{\ba}2
\mybox {\et23 \cx2{s_1}}
\Ms{\ga}1\et12
&\Rar_{EX}&\\
\cx5{s_4}\Ms{0}4\et45
\cx4{s_3}\Ms{\al}3\et34
\cx3{s_2}
\mybox{\Ms{\ba}2\cx2{s_1}}
\cz3{s_1}
\Ms{\ga}1\et1{23}
&\Rar_{MX}&\\
\cx5{s_4}\Ms{0}4\et45
\cx4{s_3}\Ms{\al}3
\mybox{\et34\cx3{s_2}\cz3{s_1}}
\ms{\ba}2{s_1}
\Ms{\ga}1\et1{23}
&\Rar_{EXZ}&\\
\cx5{s_4}\Ms{0}4\et45
\cx4{s_3}
\mybox{\Ms{\al}3\cx3{s_2}\cz3{s_1}}
\cz4{s_2}
\ms{\ba}2{s_1}
\Ms{\ga}1\et1{234}
&\Rar_{MXZ}&\\
\cx5{s_4}\Ms{0}4
\mybox{\et45\cx4{s_3}\cz4{s_2}}
\MS{\al}3{s_2}{s_1}
\ms{\ba}2{s_1}
\Ms{\ga}1\et1{234}
&\Rar_{EXZ}&\\
\cx5{s_4}
\mybox{\Ms{0}4\cx4{s_3}\cz4{s_2}}
\cz5{s_3}
\MS{\al}3{s_2}{s_1}
\ms{\ba}2{s_1}
\Ms{\ga}1\et1{2345}
&\Rar_{MXZ}&\\
\cx5{s_4}\cz5{s_3}
\MS{0}4{}{s_2}
\MS{\al}3{s_2}{s_1}
\ms{\ba}2{s_1}
\Ms{\ga}1\et1{2345}
&\Rar_{S}&\\
\cx5{s_2+s_4}\cz5{s_1+s_3}
\Ms{0}4
\MS{\al}3{s_2}{}
\ms{\ba}2{s_1}
\Ms{\ga}1\et1{2345}
}

\subsubsection*{CNOT ($\ctR X$).}
This is our first example with two inputs and two outputs.
We use here the trivial pattern $\mfr I$ with computation space $\ens1$, 
inputs $\ens1$, outputs $\ens1$, and empty command sequence,
which implements the identity over $\hil 1$.  

One has $\ctR X=(I\otimes H)\ctR Z(I\otimes H)$, so we get a pattern using 
4 qubits over $\ens{1,2,3,4}$, with inputs $\ens{1,2}$, and outputs
$\ens{1,4}$, 
where one notices that inputs and outputs intersect on the control qubit 
$\ens1$:
\AR{
(\mfr I(1)\otimes\mfr h(3,4))
\ctR\mfr Z(1,3)
(\mfr I(1)\otimes\mfr h(2,3))
&=&
\cx4{s_3}
\Ms x3
\et34
\et13
\cx3{s_2}
\Ms x2
\et23
}
By standardizing:
\AR{
\cx4{s_3}
\Ms x3
\et34
\mybox{\et13\cx3{s_2}}
\Ms x2
\et23
&\Rar_{EX}&\\
\cx4{s_3}
\cz1{s_2}
\Ms x3
\mybox{\et34\cx3{s_2}}
\Ms x2
\et13\et23
&\Rar_{EX}&\\
\cx4{s_3}
\cz4{s_2}
\cz1{s_2}
\mybox{\Ms x3\cx3{s_2}}
\Ms x2
\et13\et23\et34
&\Rar_{MX}&\\
\cx4{s_3}
\cz4{s_2}
\cz1{s_2}
\Ms x3
\Ms x2
\et13\et23\et34
}
Note that, in this case, we are not using the $\et1{234}$ abbreviation,
because the underlying structure of entanglement is not a chain.  This
pattern was already described in Aliferis and Leung's paper~\cite{AL04}.
In their original presentation the authors actually use an explicit
identity pattern (using the teleportation pattern $\mfr \G(0,0)$ presented
above), but we know from the careful presentation of composition that this
is not necessary.

\subsubsection*{GHZ.}
We present now a family of patterns preparing the GHZ entangled states
$\ket{0\ldots0}+\ket{1\ldots1}$.  One has:
\AR{
\hbox{GHZ}(n)&=&
(H_n
\ctR Z_{n-1 n}
\ldots
H_2
\ctR Z_{1 2})\ket{\hskip-.4ex+\hskip-.4ex\ldots\hskip-.4ex+}
}
and by combining the patterns for $\ctR Z$ and $H$, we obtain
a pattern with computation space $\ens{1,2,2',\ldots, n, n'}$,
no inputs, outputs $\ens{1,2',\ldots,n'}$,
and the following command sequence:
\AR{
\cx{n'}{s_n}\Ms x{n}\et{n}{n'}
\et{(n-1)'}{n}
\ldots
\cx{2'}{s_2}\Ms x{2}\et{2}{2'}
\et{1}{2}
}
With this form, the only way to run the pattern is to execute
all commands in sequence.  The situation changes completely, when we bring
the pattern to extended standard form:
\AR{
\cx{n'}{s_n}
\Ms x{n}\et{n}{n'}
\et{(n-1)'}{n}
\ldots
\cx{3'}{s_3}
\Ms x{3}
\et{3}{3'}
\mybox{\et{2'}{3}
\cx{2'}{s_2}
}\Ms x{2}\et{2}{2'}
\et{1}{2}
&\Rar&\\
\cx{n'}{s_n}
\cx{2'}{s_2}
\Ms x{n}\et{n}{n'}
\et{(n-1)'}{n}
\ldots
\cx{3'}{s_3}
\mybox{\Ms x{3}\cz{3}{s_2}}
\Ms x{2}
\et{3}{3'}
\et{2'}{3}
\et{2}{2'}
\et{1}{2}
&\Rar&\\
\cx{n'}{s_n}
\cx{2'}{s_2}
\Ms x{n}\et{n}{n'}
\et{(n-1)'}{n}
\ldots
\cx{3'}{s_3}
\MS x{3}{}{s_2}
\Ms x{2}
\et{3}{3'}
\et{2'}{3}
\et{2}{2'}
\et{1}{2}
&\Rar\st&\\
\cx{n'}{s_n}
\ldots
\cx{3'}{s_3}
\cx{2'}{s_2}
\MS x{n}{}{s_{n-1}}
\ldots
\MS x{3}{}{s_2}
\Ms x{2}
\et{n}{n'}
\et{(n-1)'}{n}
\ldots
\et{3}{3'}
\et{2'}{3}
\et{2}{2'}
\et{1}{2}
&\Rar_S&\\
\cx{n'}{s_2+s_3+\cdots+s_n}
\ldots
\cx{3'}{s_2+s_3}
\cx{2'}{s_2}
\Ms x{n}
\ldots
\Ms x{3}
\Ms x{2}
\et{n}{n'}
\et{(n-1)'}{n}
\ldots
\et{3}{3'}
\et{2'}{3}
\et{2}{2'}
\et{1}{2}
}
All measurements are now independent of each other, it is therefore
possible after the entanglement phase, to do all of them in one round, and
in a subsequent round to do all local corrections.  In other words, the
obtained pattern has constant computational depth complexity $2$.  

\subsubsection*{Controlled-$U$.}
\begin{figure}[h]
\begin{center}
\includegraphics[scale=0.6]{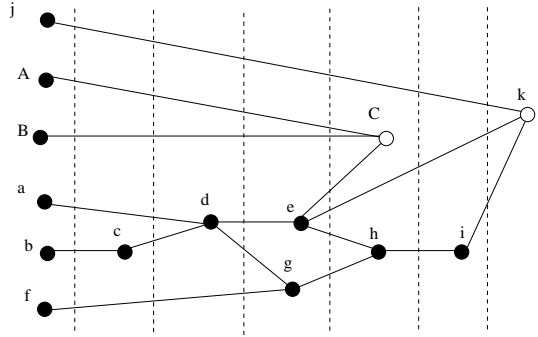}
\caption{The dependency graph for the standard $\ctR U$ pattern.}
\label{dependgraph}
\end{center}
\end{figure}

This final example presents another instance where standardization obtains
a low computational depth complexity, the proof of this fact can be found
in \cite{BK06}. 
For any 1-qubit unitary $U$, one has the following decomposition
of $\ctR U$ in terms of the generators $\G(\al)$~\cite{generator04}:
\AR{ 
\ctR U_{12}
&=& 
\Rr{1}{0}\Rr{1}{\al'}
\Rr{2}{0}\Rr{2}{\ba+\pi}
\Rr2{-\frac\ga2}\Rr2{-\pit}
\Rr20\ctR Z_{12}
\Rr2{\pit}\Rr2{\frac\ga2}
\Rr2{\frac{-\pi-\da-\ba}2} 
\Rr20
\ctR Z_{12}
\Rr2{\frac{-\ba+\da-\pi}2} 
}
with $\al'=\al+\frac{\ba+\ga+\da}2$.
By translating each $\G$ operator to its corresponding pattern, we get the
following wild pattern for $\ctR U$: 
\AR{
\cx{C}{s_B}\Ms{0}{B}\et{B}{C}
\cx{B}{s_A}\Ms{-\al'}{A}\et{A}{B}
\cx{k}{s_j}\Ms{0}{j}\et{j}{k}
\cx{j}{s_i}\Ms{-\ba-\pi}{i}\et{i}{j}
\\
\cx{i}{s_h}\Ms{\frac\ga2}{h}\et{h}{i}
\cx{h}{s_g}\Ms{\pit}{g}\et{g}{h}
\cx{g}{s_f}\Ms{0}{f}\et{f}{g}
\et{A}f
\cx{f}{s_e}\Ms{-\pit}{e}\et{e}{f}
\\
\cx{e}{s_d}\Ms{-\frac\ga2}{d}\et{d}{e}
\cx{d}{s_c}\Ms{\frac{\pi+\da+\ba}2}{c}\et{c}{d}
\cx{c}{s_b}\Ms{0}{b}\et{b}{c}
\et{A}b
\cx{b}{s_a}\Ms{\frac{\ba-\da+\pi}2}{a}\et{a}{b}
}
In order to run the wild form of the pattern one needs to follow the
pattern commands in sequence. It is easy to verify that, because of the
dependent corrections, one needs at least 12 rounds to complete the
execution of the pattern.  The situation changes completely after extended
standardization: 
\AR{
\cz{k}{s_i+s_g+s_e+s_c+s_a}
\cx{k}{s_j+s_h+s_f+s_d+s_b}
\cx{C}{s_B}
\cz{C}{s_A+s_e+s_c}
\\
\Ms{0}{B}
\Ms{-\al'}{A}
\Ms{0}{j}
\ms{-\ba-\pi}{i}{s_h+s_f+s_d+s_b}
\ms{\frac\ga2}{h}{s_g+s_e+s_c+s_a}
\ms{\pit}{g}{s_f+s_d+s_b}
\\
\Ms{0}{f}
\ms{-\pit}{e}{s_d+s_b}
\ms{-\frac\ga2}{d}{s_c+s_a}
\ms{\frac{\pi+\da+\ba}2}{c}{s_b}
\Ms{0}{b}
\Ms{\frac{\ba-\da+\pi}2}{a}
\\
\et{B}{C}\et{A}{B}
\et{j}{k}\et{i}{j}\et{h}{i}
\et{g}{h}\et{f}{g}\et{A}f\et{e}{f}\et{d}{e}\et{c}{d}\et{b}{c}\et{a}{b}\et{A}b
}
Now the order between measurements is relaxed, as one sees in 
Figure~\ref{dependgraph}, which describes the dependency structure of the
standard pattern above.  Specifically, all measurements can be completed in
$7$ rounds.  This is just one example of how standardization lowers
computational depth complexity, and reveals inherent parallelism in a
pattern.  

%%%%%%%%%%%%%%%%%%%%%%%%%%%%%%%

\section{The no dependency theorems}\label{s:expr}
From standardization we can also infer results related to dependencies.
We start with a simple observation which is a direct consequence of
standardization. 

\LE
Let $\mfr P$ be a pattern implementing some cptp-maps $T$, and suppose $\mfr
P$'s command sequence has measurements only of the $M^x$ and $M^y$ kind,
then $U$ has a standard implementation, 
having only independent measurements, all being of the $M^x$ and $M^y$ kind
(therefore of computational depth complexity at most 2).
\EL
\textbf{Proof.} Write $\mfr P'$ for the standard pattern associated to $\mfr P$.
By equations (\ref{mx}) and (\ref{my}), the $X$-actions can be eliminated
from $\mfr P'$, and then $Z$-actions can be eliminated by using the
extended calculus.  The final pattern still implements $T$, has no longer
any dependent measurements, and has therefore computational depth
complexity at most 2. 
\qed
\vskip 0.2cm

\TH \label{t-nodepend}
Let $U$ be a unitary operator, 
then $U$ is in the Clifford group iff
there exists a pattern $\mfr P$ implementing $U$, 
having measurements only of the $M^x$ and $M^y$ kind.  
\HT
\textbf{Proof.} The ``only if'' direction is easy, since we have seen in
the example section, standard patterns for $\ctR X$, $H$ and $P(\pit)$
which had only independent $M^x$ and $M^y$ measurements.  Hence any
Clifford operator can be implemented by a combination of these patterns.
By the lemma above, we know we can actually choose these patterns to be
standard. 

For the ``if'' direction, we prove that $U$ belongs to the normalizer of
the Pauli group, and hence by definition to the Clifford group.  In order
to do so we use the standard form of $\mfr P$ written as $\mfr P'= C_{\mfr
  P'}M_{\mfr P'}E_{\mfr P'}$ which still implements $U$, 
and has only $M^x$ and $M^y$ measurements.  Recall that, because of equations
(\ref{mx}) and (\ref{my}), these measurements are independent.   

Let $i$ be an input qubit, and consider the pattern $\mfr P''={\mfr
  P}'C_i$, where $C_i$ is either $X_i$ or $Z_i$.  Clearly $\mfr P''$
implements $UC_i$.   
First, one has:
\AR{
C_{\mfr P'}M_{\mfr P'}E_{\mfr P'} C_i 
&\Rar_{EC}\st& 
C_{\mfr P'}M_{\mfr P'}C' E_{\mfr P'}
}
for some \emph{non-dependent} sequence of corrections $C'$,
which, up to free commutations can be written uniquely
as $C'_OC''$, where $C'_O$ applies on output qubits, and therefore
commutes to $M_{\mfr P'}$, and $C''$ applies on non-output qubits
(which are therefore all measured in $M_{\mfr P'}$).
So, by commuting $C'_O$ both through $M_{\mfr P'}$ and $C_{\mfr P'}$ (up to
a global phase), one gets: 
\AR{
C_{\mfr P'}M_{\mfr P'}C' E_{\mfr P'}
&\Rar\st& 
C'_OC_{\mfr P'}M_{\mfr P'}C'' E_{\mfr P'}
}
Using equations (\ref{mx}), (\ref{my}), and the extended calculus
to eliminate the remaining $Z$-actions, one gets:
\AR{
M_{\mfr P'}C''
&\Rar_{MC,S}\st& 
SM_{\mfr P'}
}
for some product $S=\prod_{\ens{j\in J}}\ss j1$ of constant shifts
\footnote{Here we have used the trivial equations $Z_i^{a+1}=Z_iZ_i^a$ and
  $X_i^{a+1}=X_iX_i^a$}, applying to some subset $J$ of the non-output
qubits.  So: 
\AR{
C'_OC_{\mfr P'}M_{\mfr P'}C'' E_{\mfr P'}
&\Rar\st_{MC,S}& 
C'_OC_{\mfr P'}SM_{\mfr P'}E_{\mfr P'}\\
&\Rar\st& 
C'_OC''_OC_{\mfr P'}M_{\mfr P'}E_{\mfr P'}
}
where $C''_O$ is a further constant correction obtained
by signal shifting $C_{\mfr P'}$ with $S$.  This proves that 
$\mfr P''$ also implements $C'_OC''_OU$, and therefore 
$UC_i=C'_OC''_OU$ which completes the proof, since
$C'_OC''_O$ is a non dependent correction.  
\qed
\vskip 0.2cm

The ``only if'' part of this theorem already appears in previous
work~\cite[p.18]{mqqcs}.  The ``if'' part can be construed as an
internalization of the argument implicit in the proof of Gottesman-Knill
theorem~\cite[p.464]{NC00}.  

We can further prove that dependencies are crucial for
the universality of the model.  Observe first that if a pattern has
no measurements, and hence no dependencies, then it follows from (D2) that
$V=O$, \ie, all qubits are outputs.  Therefore computation steps involve
only $X$, $Z$ and $\ctR Z$, and it is not surprising that they compute a
unitary which is in the Clifford group.  The general argument essentially
consists in showing that when there are measurements, but still no
dependencies, then the measurements are playing no part in the result.  
\TH
Let $\mfr P$ be a pattern implementing some unitary $U$, and suppose $\mfr
P$'s command sequence doesn't have any dependencies,  
then $U$ is in the Clifford group.
\HT
\textbf{Proof.} Write $\mfr P'$ for the standard pattern associated to $\mfr P$.
Since rewriting is sound, $\mfr P'$ still implements $U$, and 
since rewriting never creates any dependency, it still has no dependencies.  
In particular, the corrections one finds at the
end of $\mfr P'$, call them $C$, bear no dependencies.
Erasing them off $\mfr P'$, results in a pattern $\mfr P''$ which is still
standard, still deterministic, and implementing $U':=C\ad U$.  

Now how does the pattern $\mfr P''$ run on some 
input $\phi$~? First $\phi\otimes\ket{\hskip-.4ex+\ldots+}$ goes by the
entanglement 
phase to some $\psi\in\hil V$, and is then subjected to a sequence of
independent 1-qubit measurements.  
Pick a basis $\mathbf{B}$ spanning the Hilbert
space generated by the non-output qubits $\hil{V\setminus O}$
and associated to this sequence of measurements.

Since $\hil V=\hil O\otimes \hil{V\setminus O}$
and $\hil{V\setminus O}=\oplus_{\phi_b\in\mcl B}[\phi_b]$,
where $[\phi_b]$  is the linear subspace generated
by $\phi_b$, by distributivity, $\psi$ uniquely 
decomposes as:
\AR{
\psi=\sum_{\phi_b\in\mcl B} x_b\otimes \phi_b
}
where $\phi_b$ ranges over $\mcl B$,
and $x_b\in\hil O$.  Now since $\mfr P''$ is deterministic, there exists an
$x$, and scalars $\la_b$ such that $x_b=\la_b x$.  Therefore $\psi$ can be
written $x\otimes \psi'$, for some $\psi'$.  
It follows in particular that the output of the computation will still be
$x$ (up to a scalar), no matter what the actual measurements are.  One can
therefore choose them to be all of the $\M x{}$ kind, 
and by the preceding theorem $U'$ is in the Clifford group, and so is
$U=CU'$, since $C$ is a Pauli operator.  
\qed
\vskip 0.2cm

From this section, we conclude in particular that any universal set of patterns
has to include dependencies (by the preceding theorem), and also needs
to use measurements $M^\al$ where $\al\neq0$ modulo $\pit$ (by the theorem
before).  This is indeed the case for the universal set $\mfr\G(\al)$ and
$\ctR{\mfr Z}$.  

%%%%%%%%%%%%%%%%%%%%%%%%%%%%%%%

\section{Other Models}\label{s:other}
There are several other approaches to measurement-based computation as we
have mentioned in the introduction.  However, it is only for the one-way
model that the importance of having all the entanglement in front has been
emphasized.  For example, Gottesman and Chuang describe computing with
teleportation in the setting of the circuit model and hence the computation
is very sequential \cite{Gott99}.  What we will do is to give a general
treatment of a variety of measurement-based models -- including some that
appear here for the first time -- in the setting of our calculus.  More
precisely we would like to know other potential definitions for commands
$N$, $E$, $M$ and $C$ that lead to a model that still satisfies the
properties of: (i) being closed under composition; (ii) universality and
(iii) standardization.

Moreover we are interested in obtaining a compositional embedding of these
models into a single \emph{one-qubit} measurement-based model.  The
teleportation model can indeed be embedded into the one-way model.  There
is, however, a new model, the Pauli model -- formally defined here for the
first time -- which is motivated by considerations of fault tolerance
\cite{RAB04,dk05a,DKOS06}.  The Pauli model can be embedded into a slight
generalization of the one-way model called the phase model; also given here
for the first time.  The one-way model will trivially embed in the phase
model so by composition all the measurement-based models will embed in the
phase model.  We could have done everything \emph{ab initio} in terms of
the phase model but this would have made much of the presentation
unnecessarily complicated at the outset.

We recall the remark from the introduction that these embeddings have three
advantages: first, we get a workable syntax for handling the dependencies
of operators on previous measurement outcomes, second, one can use these
embeddings to transfer the measurement calculus previously developed for
the one-way model to obtain a calculus for the new model including, of
course, a standardization procedure that we get automatically; lastly, one
can embed the patterns from the phase model into the new models and vice
versa.  In essence, these compositional embeddings will allow us to exhibit
the phase model as being a core calculus for measurement-based computation.
However different models are interesting from the point of view of
implementation issues like fault-tolerance and ease of preparation of
entanglement resources.  Our embeddings allow one to move easily
between these models and to concentrate on the one-way model for designing
algorithms and proving general theorems.

This section has been structured into several subsections, one for each
model and its embedding.  

\subsection{Phase Model}
In the one-way model the auxiliary qubits are initialized to be in the
$\ket{+}$ state.  We extend the one-way model to allow the auxiliary qubits
to be in a more general state.  We define the extended preparation command
$\Nq i\al$ to be the preparation of the auxiliary qubit $i$ in the state
$\ket{+_{\al}}$.  We also add a new correction command $\Cp i \al$, called
a \emph{phase correction} to guarantee that we can obtain determinate
patterns.  
The dependent phase correction is written as $\cp i\al s$ with
$\cp i \al 0 = I$ and $\cp i \al 1= \MA{1&0\\0&\ei\al}$.  Under
conjugation, the phase correction, defines a new action over measurement:
\AR { ( \cp i \ba s ){}^{\dag} \M{{\al}}i \cp i \ba s=\M {\al-s\ba}
  i\label{mp} } 
and since the measurement is destructive, it simplifies to $\M{\al}i \cp i
\ba s =\M {\al-s\ba}i$.  This action does not commute with Pauli actions
and hence one cannot write a compact notation for dependent measurement, as
we did before, and the computation of angle dependencies is a bit more
complicated.  Thereafter, a measurement preceded by a sequence of
corrections on the same qubit will be called a dependent measurement.  Note
that, by the absorption equations, this indeed can be seen as a
measurement, where the angle depends on the outcomes of some other
measurements made beforehand.

To complete the extended calculus it remains to define the new rewrite
rules:
\AR
{  
\et ij \cp i \al s &\Rar& \cp i \al s \et ij &\quad\hbox{}EP\\\\

\M \al i \cx i s &\Rar&\M {(-1)^s\al} i&\quad\hbox{}MX\\
\M \al i \cz i s  &\Rar&\M {\al-s\pi} i  &\quad\hbox{}MZ\\
\M \al i \cp i\ba s &\Rar&\M {\al-s\ba} i &\quad\hbox{}MP } The above rules
together with the rewriting rules of the one-way model described in Section
\ref{s-cal}, lead to a standardization procedure for the model.  It is
trivial that the one-way model is a fragment of this generalized model and
hence universality immediately follows.  It is also easy to check that the
model is closed under composition and all the semantical properties of the
one-way model can be extended to this general model as well.  

The choice of extended preparations and its concomitant phase correction is
actually quite delicate.  One wishes to keep the standardizability of the
calculus which constrains what can be added but one also wishes to have
determinate patterns which forces us to put in appropriate corrections.
The phase model is only a slight extension of the original one-way model,
but it allows a discussion of the next model which is of great physical
interest.

\subsubsection{Pauli model}
An interesting fragment of the phase model is defined by restricting the
angles of measurements to $\ens{0,\pit,\pi,-\pit}$ \ie \, Pauli
measurements and the angles of preparation to $0$ and $\pif$.  Also the
correction commands are restricted to Pauli corrections $X$, $Z$ and Phase
correction $Z^{\frac{\pi}{8}}$.  One readily sees that the subset of angles
is closed under the actions of the corrections and hence the Pauli model is
closed under composition.

\PRO
The Pauli model is approximately universal.
\ORP
\textbf{Proof.}
We know that the set consisting of $\G(0)$ (which is $H$), $\G({\pif})$,
and $\ctR Z$ is approximately universal.  Hence, to prove the approximate
universality of Pauli model, it is enough to exhibit a pattern in the Pauli
model for each of 
these three unitaries.  We saw before that $\G(0)$ and $\ctR Z$ are
computed by the following 2-qubit patterns: \AR{
  \mfr \G(0)&:=&\cx 2{s_1}\M 0 1\et 12\\
  \ctR{\mfr Z}&:=&\et 12 } where both belong also to the Pauli model.
The pattern for $\G_{\pif}$ in the one-way model is
expressed as follows: \AR{
  \mfr \G({\pif})&:=&\cx 2{s_1}\M {-\pif}1\et 12\\
  &=&\cx 2{s_1}\M 0 1\et 12\Cp 1{\pif} } The above forms do not fit in the
Pauli model, since the first one uses a measurement with an angle $\pif$
and the second uses $\Cp {}{\pif}$.  However by teleporting the input qubit
and then applying the $\Cp {}{\pif}$ and finally running the
standardization procedure we obtain the following pattern in the Pauli
model for $J(\pif)$: \AR{ 
  & &\cx 2{s_1}\M{0}1\et 12\Cp 1\pif\\
  &=&  \cx 4{s_3}\M{0}3\et 34 \mybox{\Cp 3\pif} \cz 3 {s_2} \cx 3
  {s_1}\M{0}2\M{0}1\et 12\et23\\ 
  &=&  \cx 4{s_3}\M{0}3\et 34\cz 3 {s_2}\cp 3\pit {s_2}\cx 3
  {s_1}\M{0}2\M{0}1\et 12\et23\mybox{\Cp 3\pif}\\ 
  &=&  \cx 4{s_3}\M{0}3\et 34\cz 3 {s_2}\cp 3\pit {s_2}\cx 3
  {s_1}\M{0}2\M{0}1\et 12\et23\mybox{\Cp 3\pif}\\ 
  &=& \cx 4{s_3+s_2} \cz 4{s_1} \M{-(-1)^{s_1}{s_2}\pit}3 \M{0}2\M{0}1\et
  12\et23 \et 34N_3^{\pif} }
Approximate universality for the Pauli model is now immediate.  \qed
\vskip 0.2cm

Note that we cannot really expect universality (as we had for the phase
model) because the angles are restricted to a discrete set.  On the other
hand it is precisely this restriction that makes the Pauli model
interesting from the point of view of implementation.  The other particular
interest behind this model, apart from its simple structure, is based on
the existence of a novel fault tolerant technique for computing within this
framework \cite{Bra04,RAB04,DKOS06}.

\subsection{Teleportation}
Another class of measurement-based models -- older, in fact, than the
one-way model -- uses $2$-qubit measurements.  These are collectively
referred to as \emph{teleportation models}~\cite{L03}.  Several papers that
are concerned with the relation and possible unification of these
models~\cite{CLN04,AL04,JP04} have already appeared.  One aspect of these
models that stands in the way of a complete understanding of this relation,
is that, whereas in the one-way model one has a clearly identified class of
measurements, there is less agreement concerning which measurements are
allowed in teleportation models.

We propose here to take as our class of $2$-qubit measurements a family
obtained as the conjugate under the operator $\ctR Z$ of tensors of
$1$-qubit measurements.  We show that the resulting teleportation model is
universal.  Moreover, almost by construction, it embeds into the one-way
model, and thus exposes completely the relation between the two models.

Before embarking on the specifics of our family of $2$-qubit measurements,
we remark that the situation commented above is more general:
\LE\label{2to1m} Let $\cal{A}$ be an orthonormal basis in $\otimes^n\ctwo$,
with associated $n$-qubit measurement $M^{\mcl{A}}$, and $\mcl{A}_i$ with
$i=1,\ldots,n$ be orthonormal bases in $\ctwo$, with associated $1$-qubit
measurements $M^{\mcl{A}_i}_i$.  Then there exists a unique (up to a
permutation) $n$-qubit unitary operator $U$ such that: 
\AR{
M^{\mcl{A}}_{1\cdots n}&=& U_{1\cdots n} (\otimes_i
M^{\mcl{A}_i}_i)U\st_{1\cdots n}  
} 
\EL 
\textbf{Proof.} Take $U$ to map $\otimes_i \mcl A_i$ to $\mcl A$.  \qed
\vskip 0.2cm

This simple lemma says that general $n$-qubit measurements can always be
seen as conjugated $1$-qubit measurements, provided one uses the
appropriate unitary to do so.  As an example consider the orthogonal
\emph{graph basis} $\mcl G=\ctR Z_{12}\ens{\ket\pm\otimes\ket\pm}$ then the
two-qubit graph basis measurements are defined as $M^{\mcl G}_{12}= \ctR
Z_{12} (\M 0 1\otimes \M 0 2) \ctR Z_{12}$.  It is now natural to extend
our definition of $M^{\mcl G}_{12}$ to obtain the family of $2$-qubit
measurements of interest: \EQ{\label{e2qubit} \tM \al\ba 12 &:=& \ctR
  Z_{12}(\M \al1\otimes\m\ba2)\ctR Z_{12} } corresponding to projections on
the basis $\mcl G_{\al,\ba}:= \ctR Z_{12}(P_1(\al)\otimes
P_2(\ba))(\ens{\ket{\pm}\otimes\ket{\pm}}$.  This family of two-qubit
measurements together with the preparation, entanglement and corrections
commands of the one-way model define the teleportation model.

Before we carry on, a clarification about
our choice of measurements in the teleportation model is necessary.  The
usual teleportation protocol uses Bell basis measurement defined with
\AR{
\mcl B&=&\ctR X_{12}\ens{\ket\pm\otimes\ket{0/1}} \\
M^{\mcl B}_{12}&=& \ctR X_{12} (\M z 1\otimes \M z 2) \ctR X_{12}
}
where $M^z$ is the computational-basis measurement.  Note how similar these
equations are to the equations defining the graph basis measurements.  This
is a clear indication that everything that follows can be transferred to
the case where $X$ replaces $Z$, and $\mcl B$ replaces $\mcl G$.  However,
since the methodology we adopt is to embed the 2-qubit measurement based
model in the one-way model, and the latter is based on $\ctR Z$ and $\mcl
G$, we will work with the graph-basis measurements.  Furthermore, since
$\ctR Z$ is symmetric, whereas $\ctR X$ (\aka\ as C-NOT) is not, the
algebra is usually nicer to work with.

Now we prove that the family of measurements in Equation~\ref{e2qubit}
leads to a universal model, which embeds nicely into the one-way model, but
first we need to describe the important notion of dependent measurements.
These will arise as a consequence of standardization; they were not
considered in the existing teleportation models.

In what follows we drop the subscripts on the $\ctR Z$ unless they are
really necessary.  We write $(s(i),s(j))\in\ztwo\times\ztwo$ to represent
outcome of a 2-qubit measurement, with the specific convention that
$(0,0)$, $(0,1)$, $(1,0)$, and $(1,1)$, correspond respectively to the
cases where the state collapses to $\ctR Z\ket{+_\al}\ket{+_\al}$, $\ctR
Z\ket{+_\al}\ket{-_\al}$, $\ctR Z\ket{-_\al}\ket{+_\al}$, and $\ctR
Z\ket{-_\al}\ket{-_\al}$.

We will use two types of dependencies for measurements
associated with $X$-action and $Z$-action:
\AR{
\TMR\al\ba ijst
&=&\tM {(-1)^s\al}{(-1)^t\ba} ij
\\
\TML\al\ba ijuv
&=&\tM {\al+u\pi}{\ba+v\pi} ij
}
where $s$, $t$, $u$ and $v$ are in $\ztwo$.  The two actions commute, so the
equations above  
define unambiguously the full dependent measurement $\TMLR \al\ba
ijstuv$.  Here are some useful abbreviations: 
\AR{
\TMLR\al\ba {}{}st00:=\TMR\al\ba {}{}st\\
\TMLR\al\ba {}{}00uv:=\TML\al\ba {}{}uv\\
\TMLR\al\ba {}{}0000:=\tM \al\ba {}{}\\
\tM \al x {}{}:=\tM \al0 {}{}\\
\tM \al y {}{}:=\tM \al\pit {}{}\\
}

As in the 1-qubit measurement case we obtain the following rewriting
rules for the teleportation model: 
\AR{
\et ij\cx is&\Rar&\cx is\cz js\et ij&\quad\hbox{}EX\\
\et ij\cz is&\Rar&\cz is\et ij&\quad\hbox{}EZ\\
\TMLR\al \ba  ijstuv\cx ir&\Rar&\TMLR \al \ba ij{s+r}tu{v+r}&\quad\hbox{}MX\\
\TMLR\al \ba  ijstuv\cx jr&\Rar&\TMLR \al \ba ijs{t+r}{u+r}v&\quad\hbox{}MX\\
\TMLR\al \ba  ijstuv\cz ir&\Rar&\TMLR \al \ba ijst{u+r}v&\quad\hbox{}MZ\\
\TMLR\al \ba  ijstuv\cz jr&\Rar&\TMLR \al \ba ijstu{v+r}&\quad\hbox{}MZ
}
to which we add also the trivial commutation rewriting which are possible
between commands that don't overlap (meaning, acting on disjoint sets of
qubits).  

\subsubsection{Embedding}

We describe how to translate $2$-qubit EMC patterns to $1$-qubit patterns and
vice versa.  The following equation plays the central role in the
translation: \EQ{\label{s2to1m} \tM \al\ba ij=\et ij(\m \al i\otimes\m \ba
  j)\et ij }
Note that this immediately gives the denotational semantics of two-qubit
measurements as cptp-maps.  Furthermore, all other commands in the
teleportation model are the same as in the one-way model, so we have right
away a denotational semantics for the entire teleportation model in terms of
cptp-maps.  

We write $\mathfrak{P}$ for the collection of patterns in the
one-way model and $\mathfrak{T}$ for the collection of
patterns in the teleportation model.
\begin{theo}
There exist functions $[\cdot]_f:\mathcal{P}\to\mathcal{T}$ and 
$[\cdot]_b:\mathcal{T}\to\mathcal{P}$ such that
\begin{enumerate}
\item $\forall \mfr{P}\in\mathfrak{P}:\sem{\mfr P} =
  \sem{[\mfr P]_f}$;
\item $\forall \mfr T\in\mathfrak{T}:\sem{\mfr T} =
  \sem{[\mfr T]_b}$;
\item $[\cdot]_f\circ[\cdot]_b$ and $[\cdot]_b\circ[\cdot]_f$ are both
  identity maps.
\end{enumerate}
\end{theo}
\textbf{Proof.}
We first define the forward map $[\cdot]_f$ in stages as follows for any
patterns $\mfr P = (V,I,O,A_n\ldots A_1)$:
\begin{enumerate}
\item For any $i\in V\setminus O$ (i.e.\ measured qubits) we add an
  auxiliary qubit $i_d$ called a \emph{dummy} qubit to the space $V$.
\item For any $i\in V\setminus O$ we replace any
  occurrence of $M_i^{\alpha}$ with $M_i^{\alpha}M_{i_d}^x$\,.
\item We then replace each of the newly created occurrences of
  $M_i^{\alpha}M_{i_d}^x$ by $M_{ii_d}^{\alpha,x}E_{ii_d}$\,.
\end{enumerate}
Now we show that the first condition stated in the theorem holds; we do
this stage wise.  The first two stages are clear because we are just adding
qubits that have no effect on the pattern because they are not entangled
with any pre-existing qubit, and no other command depends on a measurement
applied to one of the dummy qubits.  Furthermore, we add qubits in the
state $\ket{+}$ and measure them in the $\ket{\pm}$ basis.  The invariance of
the semantics under stage 3 is an immediate consequence of
Equation~\ref{s2to1m} and the fact that all the measurements are
destructive, and hence an entanglement command on qubits appearing after a
measurement of any of those qubits can just be removed.

The map $[\cdot]_b$ is defined similarly except that there is no need to
add dummy qubits.  One only needs to replace any two-qubit measurement
$M^{\alpha,\beta}_{ij}$ with $M_i^{\alpha}M_j^{\beta}E_{ij}$.  Again, this
clearly preserves the semantics of patterns because of
Equation~\ref{s2to1m} and the above remark about destructive measurements.
Thus condition 2 of the theorem holds.

The fact that the two maps are mutual inverses follows easily.
As all the steps in the translations are local we
can reason locally.  Looking at the forward mapping followed by the
backward mapping we get the following sequence of transformations
\AR
{M_i^{\alpha} & \Rightarrow_{\text{stage }1,2} &
M_i^{\alpha}M_{i_d}^x\\ & \Rightarrow_{\text{Equation \ref{s2to1m}}} &
M_{ii_d}^{\alpha,x} E_{ii_d}\\ & \Rightarrow_{\text{Equation \ref{s2to1m}}} &
M_i^{\alpha}M_{i_d}^x E_{ii_d}E_{ii_d}\\ & \Rightarrow &
M_i^{\alpha}M_{i_d}^x\\ & \Rightarrow &
M_i^{\alpha} 
}
This shows that we have the third condition of the theorem.
\qed
\vskip 0.2cm

Note that the translations are compositional since the denotational
semantics is and also it follows immediately that the teleportation model
is universal and admits a standardization procedure.

\emph{Example.  } Consider the teleportation pattern in the teleportation
model given by the command sequence: $\cx 3{s_1} \cz 3{s_2}\tM xx 12 \et
23$, we perform the above steps:   
\AR{
\cx 3{s_1} \cz 3{s_2}\mybox{\tM xx 12} \et 23
&\Rar_{\text{Equation \ref{s2to1m}}}&\\
\cx 3{s_1} \cz 3{s_2} \m x 1\m x 2 \et 12\et 23
}   
and hence obtain the teleportation pattern with $1$-qubit measurements.

\vskip 0.5cm

\emph{Example.}  We saw before, the following EMC $1$-qubit pattern for
$R_{z}(\al)$ which can be embedded to an EMC $2$-qubit pattern using the
above steps: \AR{ \cx3{s_2} \cz3{s_1} \mybox{\mR x2{s_1} \m{-\al}1}
  \et12\et23
  &\Rar_{\text{stage 1,2}}&\\
  \cx3{s_2} \cz3{s_1} \mybox{\mR x2{s_1} \M x{2_d} \m{-\al}1 \M x{1_d}}
  \et12\et23
  &\Rar_{\text{Equation \ref{s2to1m} and standardization}}&\\
  \cx3{s_2} \cz3{s_1} \TMR x x 2{2_d}{s_1}0 \tM {-\al} x 1{1_d} \et
  1{1_d}\et 2 {2_d}\et12\et23 }
  
Note that we have explicit algorithmic translations between the models and
not just illustrative examples.  This is the main advantage of our approach
in unifying these two models compared to the extant work \cite{CLN04,AL04,JP04}.

%%%%%%%%%%%%%%%%%%%%%%%%%%%%%%%

\section{Conclusion}\label{s:conc}
We have presented a calculus for the one-way quantum computer.  We have
developed a syntax of patterns and, much more important, an \emph{algebra}
of pattern composition.  We have seen that pattern composition allows for a
structured proof of universality, which also results in parsimonious
implementations.  We develop an operational and denotational semantics for
this model; in this simple first-order setting their equivalence is clear.

We have developed a rewrite system for patterns which preserves the
semantics.  We have shown further that our calculus defines a
polynomial-time standardization algorithm transforming any pattern to a
standard form where entanglement is done first, then measurements, then
local corrections.  We have inferred from this procedure that the
denotational semantics of any pattern is a cptp-map and also proved that
patterns with no dependencies, or using only Pauli measurements, may only
implement unitaries in the Clifford group.

In addition we introduced some variations of the one-way and teleportation
models and presented compositional back-and-forth embeddings of these
models into the one-way model.  This allows one to carry forward all the
theory we have developed: semantics, rewrite rules, standardization,
no-dependency theorems and universality.  In fact the result of making the
connection between the one-way model and the teleportation model is to
introduce ideas: dependent measurements, standard forms for patterns and a
standardization procedure which had never been considered before for the
teleportation model.  This shows the generality of our formalism: we expect
that any yet to be discovered measurement-based computation frameworks can
be treated in the same way.

Perhaps the most important aspect of standardization is the fact that now
we can make patterns maximally parallel and distributed because all the
entanglement operators, \ie~ non-local operators, can be performed at the
beginning of the computation.  Then from the dependency structure that can
be obtained from the standard form of a pattern the measurements can be
organized to be as parallel as possible.  This is the essence of the
difference between measurement-based computation and the quantum circuit
model or the quantum Turing machine.

We feel that our measurement calculus has shown the power of the formalisms
developed by the programming languages community to analyze quantum
computations.  The ideas that we use: rewriting theory, (primitive) type
theory and above all, the importance of reasoning compositionally, locally
and modularly, are standard for the analysis of traditional programming
languages.  However, for quantum computation these ideas are in their
infancy.  It is not merely a question of adapting syntax to the quantum
setting; there are fundamental new ideas that need to be confronted.  What
we have done here is to develop such a theory in a new,
physically-motivated setting.

There were prior discussions about putting patterns in a standard
form~\cite{RB02} but these worked only with strongly deterministic patterns,
furthermore one needs to know which unitary is being implemented.  In our
case the rewrite rules are entirely local and work equally well with all
patterns.

An interesting question related to the measurement calculus is whether one
can give sufficient conditions -- depending only on the entanglement
structure of a pattern -- that guarantee determinacy.  In a related paper
the first two authors have solved this problem~\cite{dk05c}.  In effect
given an entanglement structure with distinguished inputs and outputs one
can enumerate all the unitaries that can be implemented with it.  This
gives a precise handle on the entanglement resources needed in the design
of specific algorithms and protocols directly in the measurement-based
model~\cite{BDK06}.

Finally, there is also a compelling reading of dependencies as classical
communications, while local corrections can be thought of as local quantum
operations in a multipartite scenario.  From this point of view,
standardization pushes non-local operations to the beginning of a
distributed computation, and it seems the measurement calculus could prove
useful in the analysis of distributed quantum protocols.  To push this idea
further, one needs first to articulate a definition of a distributed
version of the measurement calculus; this was done in a recent
paper~\cite{dhkp05}.  The distributed version of the calculus was then used
to analyze a variety of quantum protocols and to examine the notion of
knowledge flow in them~\cite{DP05}.

%%%%%%%%%%%%%%%%%%%%%%%%%%%%%%%

\section*{Acknowledgments}
We thank the anonymous referees for their helpful comments.  We would like
to thank Ellie D'Hondt for implementing an interpreter for the measurement
calculus and Daniel Gottesman for clarifying the extent to which
Theorem~\ref{t-nodepend} was implicit in his own work.  We have benefited
from discussions with Samson Abramsky, Hans Briegel, Dan Browne, Philippe
Jorrand, Harold Olivier, Simon Perdrix and Marcus Silva.  Elham Kashefi was
partially supported by the PREA, MITACS, ORDCF, CFI and ARDA projects
during her stay at University of Waterloo where this work was finished.
Prakash Panangaden thanks EPSRC and NSERC for support and Samson Abramsky
and the Oxford University Computing Laboratory for hospitality at Oxford
where this work was begun.

%%%%%%%%%%%%%%%%%%%%%%%%%%%%%%%

\newcommand{\etalchar}[1]{$^{#1}$}

%%%%%%%%%%%%%%%%%%%%%%%%%%%%%%%

\appendix
\newcommand{\pair}[2]{\mbox{$\langle #1,#2\rangle$}}
\newcommand{\ketbra}[2]{| #1 \rangle\langle #2 |}
\newcommand{\braket}[2]{\langle #1 | #2 \rangle }
\newcommand{\adj}[1]{#1^{\dagger}}
\newcommand{\hilb}{\mathfrak H}
\newcommand{\comp}{\mathbb{C}}
\newcommand{\proj}[1]{| #1 \rangle\langle #1 |}

\section{Background on Quantum Mechanics and Quantum 
Computation} 

We give a brief summary of quantum mechanics and quantum computing.
We develop some of the algebra, define some notations, and prove a couple
of equations which we have used in the paper.  Although the paper is
self-contained, the reader will find the expository book of Nielsen and
Chuang~\cite{NC00} useful for quantum computation or the excellent book by
Peres~\cite{Peres95} for general background on quantum mechanics.

\subsection{Linear Algebra for Quantum Mechanics}

We assume that the reader is familiar with the basic notion of a vector
space.  In quantum mechanics we always consider vector spaces over the
complex numbers.  For quantum computation the vector spaces are always
finite dimensional.  The vector spaces that arise in quantum mechanics are
\emph{Hilbert spaces} and are thus usually written $\hilb$; that is they
have an inner product usually written $\pair{u}{v}$ where $u$ and $v$ are
vectors.  The inner product is a map from $\hilb\times\hilb$ to the complex
numbers $\comp$.  The inner product is linear in the second argument but
\emph{anti}-linear in the first argument.  In general, there is a
topological completeness condition on Hilbert spaces but, in the finite
dimensional case this is automatic and we will ignore it.

Following Dirac, it is customary to call elements of $\hilb$ \emph{kets}
and write them in the form $\ket{u}$ or whatever symbol is appropriate
inside the half-bracket.  The dual vectors are called \emph{bras} and are
written $\bra{v}$; the pairing thus can naturally be identified --
conceptually and notationally -- with the inner product.

Linear operators come naturally with vector spaces; a \emph{linear
  operator} is a linear map from a vector space to itself.  Linear
operators on finite dimensional spaces are often represented as matrices.
The most important notion for an operator on a Hilbert space is that of an
\emph{adjoint}.  
\DE 
If $A:\hilb\to\hilb'$ is a linear operator then the adjoint, written
  $\adj{A}$, is a linear operator from $\hilb'$ to $\hilb$ such that
\[ \forall u\in\hilb',v\in\hilb \pair{u}{Av} = \pair{\adj{A}u}{v}.\]
\ED

In terms of matrices this just amounts to transposing the matrix and complex
conjugation each of the matrix entries; sometimes this is called the
hermitian conjugate.  An inner product preserving linear map is called 
a \emph{unitary embedding}.  When $\hilb=\hilb'$ we can also define 
the following operators.  A \emph{hermitian operator} $A$ is one such that $A =
A^{\dagger}$ and a \emph{unitary operator} $U$ is one such that $U^{-1} =
U^{\dagger}$.  A \emph{projection} $P$ is a linear operator such that $P^2
= P$ and $P = P^{\dagger}$.  A projection operator can be identified with a
subspace, namely its range.  The eigenvalues of a hermitian operator are
always real.  Suppose $U$ is a unitary, and $P$ a projection, then
$UPU^\dag$ is also a projection.

It is common to use the Dirac notation to write projection operators as
follows: given a vector $\ket{a}$ of unit norm, the projection onto the
subspace spanned by $\ket{a}$ is written $\proj{a}$.  To see why this makes
sense, suppose that $\ket{b}$ is another vector then its component along
$\ket{a}$ is the inner product $\pair{a}{b}$.  Now if we just juxtapose the
expressions $\proj{a}$ and $\ket{b}$ we get $\ket{a}\pair{a}{b}$, viewing
the $\pair{a}{b}$ as a number and moving it to the front we get
$\pair{a}{b}\ket{a}$ as the result, which is the right answer for the
projection of $\ket{b}$ onto $\ket{a}$.  Thus one can apply the projection
operator just by juxtaposing it with the vector.  This kind of suggestive
manipulation is part of the appeal of the Dirac notation.

One important fact -- the spectral theorem for hermitian operators -- states
that if $M$ is a hermitian operator, $\lambda_i$ are its eigenvalues and
$P_i$ are projection operators onto the corresponding eigenspaces then one
can write 
\[ M = \sum_i \lambda_i P_i.\]
If we have $\ket{i}$ as the normalized eigenvectors for the eigenvalues
$\lambda_i$ then we can write this in Dirac notation as:
\[ M = \sum_i \lambda_i \proj{i}.\]

Finally we need to combine Hilbert spaces.  
\DE
Given two Hilbert spaces
$\hilb$ with basis vectors $\{a_i| 1\leq i\leq n\}$ and $\hilb'$ with basis
$\{b_j| 1\leq j\leq m\}$ we define the \emph{tensor product}, written
$\hilb\otimes\hilb'$, as the vector space of dimension $n\cdot m$ with
basis $a_i\otimes b_j$.
\ED
There are more elegant, basis-independent ways of describing the tensor
product but this definition will serve our needs.  We almost never write
the symbol $\otimes$ between the vectors.  In the Dirac notation this is
always omitted and one writes, for example, $\ket{uv}$ instead of
$\ket{u}\otimes\ket{v}$.  

The important point is that there are vectors that cannot be written as the
tensor product of vectors.  For example, we can write $a_1\otimes b_1 +
a_2\otimes b_2$ where the $a_i$ and the $b_i$ are basis vectors of two
$2$-dimensional Hilbert spaces.  This means that given a general element of
$\hilb\otimes\hilb'$ one cannot produce elements of $\hilb$ and $\hilb'$;
this is very different from the cartesian product of sets.  This is the
mathematical manifestation of entanglement.

A very important function on square matrices is the \emph{trace}.  The
usual trace -- i.e.\ the sum of the diagonal entries -- is basis independent
and is actually equal to the sum of the eigenvalues, counted with
multiplicity.  The trace of $A$ is written $tr(A)$ and satisfies the
cyclicity property $tr(AB) = tr(BA)$; applying this repeatedly one gets
\[ tr(A_1\ldots A_n) = tr(A_{\sigma(1)}\ldots A_{\sigma(n)}) \] where
$\sigma$ is a cyclic permutation.  The explicit formula for the trace of
$A:V\to V$ is $tr(A) = \sum_i \bra{i}A\ket{i}$ where $\ket{i}$ is a basis
for $V$.  

One often needs to compute a \emph{partial} trace.  Consider a linear map 
$L:V\otimes W\to V\otimes W$.  Suppose that $\ket{v_i}$ is a basis for $V$
and $\ket{w_i}$ is a basis for $W$ then $\ket{v_i w_j}$ is a basis for
$V\otimes W$.  Now we can define the partial trace over $V$ as
\[ tr_V(A):W\to W = \sum_i \bra{v_i}A\ket{v_i}.\]
This corresponds to removing the $V$ dependency; often we use the phrase 
``tracing out the $V$ component.''

\subsection{Quantum Mechanics}
We state the basic facts of quantum mechanics and will not discuss the
experimental basis for this framework.
The key aspects of quantum mechanics are:
\begin{itemize}
\item the states of a quantum system form a Hilbert space,
\item when two quantum systems are combined, the state space of the
  composite system is obtained as the tensor product of the state spaces of
  the individual systems, and
\item the evolution of a quantum system is given by a unitary operator, and
\item the effect of a measurement is indeterminate.
\end{itemize}

The first says that one can form \emph{superpositions} of the states.  This
is one of the most striking features of quantum mechanics.  Thus states are
not completely distinct as they are in classical systems.  The inner
product measures the extent to which states are distinct.  The fact that
systems are combined by tensor product says that there are states that
of composite systems that cannot be decomposed into individual pieces.
This is the phenomenon of entanglement or \emph{non-locality}.

Measurement is what gives quantum mechanics its indeterminate character.
The usual case, called projective measurements, is when the quantity being
measured is described by a hermitian operator $M$.  The possible outcomes
are the \emph{eigenvalues} of $M$.  If $M$ is an observable (hermitian
operator) with eigenvalues $\lambda_i$ and eigenvectors $\ket{\phi_i}$ and
we have a generic state $\ket{\psi} = \sum_i c_i\ket{\phi_i}$ then the
probabilities and expectation values of the measurement outcomes are given
by:
\begin{itemize}
\item $Prob(\lambda_i|\ket{\psi}) = |c_i|^2$
\item $E[M|\ket{\psi}] = \sum_i |c_i|^2 \lambda_i = \sum_i
  c_i\bar{c_i}\pair{\phi_i}{M\phi_i} = \pair{\psi}{M\psi}.$
\end{itemize}
It is important to note that the effect of the measurement is that the
projection operator $P_i$ is applied when the result $\lambda_i$ is
observed.  The operator $M$ does not describe the effect of the
measurement.  

The formulas above reveal that no aspect of a measurement is altered if the
vector describing a quantum state is multiplied by a complex number of
absolute value $1$.  Thus we can multiply a state by $e^{i\theta}$ without
changing the state.  This is called changing the phase.  While the phase is
not important phase differences are very important.  Multiplying a vector
by a complex number is a change of phase as well as a change in its
length.  Usually we normalize the state so that we can read the results of
measurements as probabilities.  Given a vector the subspace spanned by it -
always one dimensional -- is called a \emph{ray}.  Thus a state is really a
ray rather than a vector.  However, it is customary to blur this
distinction.  

\subsection{Some qubit algebra}
Quantum computation is carried out with \emph{qubits} the quantum analogues
of bits.  Just as a bit has two possible values, a qubit is a two
dimensional complex Hilbert space, in other words it is (isomorphic to) the
two dimensional complex vector space $\ctwo$.

One works with a preferred basis, physically this corresponds to two
distinguishable states, like ``spin up'' and ``spin down''.
One writes $\ket0$, and
$\ket1$ for its canonical basis, so that any vector $\psi$ can be written
as $\al \ket0+ \ba\ket1$ with $\al$, $\ba$ in $\mbb C$.  Furthermore,
$\ctwo$ can be turned into a Hilbert space with the following inner
product: 
\AR{ \tuple{\al\ket0+\ba\ket1,\al'\ket0+\ba'\ket1}&:=&\al\st
  \al'+\ba\st \ba' } where $\al\st$ is the complex conjugate of $\al$.  One
then obtains the norm of a vector as: \AR{
  \norm{\psi}&:=&\tuple{\psi,\psi}^{\frac12}&=&(\al\st \al+\ba\st
  \ba)^{\frac12} }

Given $V$ a finite set, one writes $\hil V$ for the Hilbert space
$\otimes_{u\in V}\ctwo$; the notation means an $n$-fold tensor product of
the $\ctwo$ where $n$ is the size of $V$.  A vector in $\hil V$ is said to
be \emph{decomposable} if it can be written $\otimes_{u\in V}\psi_u$ for
some $\psi_u\in\ctwo$.  Such decomposable vectors will be written $\eps$ in
the sequel.  Decomposable vectors can be represented by a
map from $V$ to $\ctwo$, and we will use both notations depending on which
is more convenient.  As we have noted before there are some vectors that
are not decomposable.

As in the case of $\ctwo$, there is a canonical basis for $\hil V$,
sometimes also called the \emph{computational basis}, containing
decomposable vectors $\eps$ such that for all $v\in V$, $\eps(v)=\ket0$ or
$\eps(v)=\ket1$.

The inner product on $\hil V$, according to the general definition given
above, is defined on decomposable vectors as: 
\AR{
\tup{\eps,\eps'}&:=&\prod_{v\in V}\tup{\eps(v),\eps'(v)}
}
Note that all vectors in the computational basis are orthogonal and of norm
$1$.  The vectors of norm $1$ are usually called \emph{unit vectors}; we
always assume that states are described by unit vectors as noted before.

Here are some common states that arise in quantum computation:
\[ \ket{0} = \ket{\uparrow} = \left[ \begin{array}{c} 1 \\ 0
  \end{array}\right], 
   \ket{1} = \ket{\downarrow} = \left[ \begin{array}{c} 0 \\ 1 
  \end{array}\right],  
   \ket{+} = \frac{1}{\sqrt{2}} \left[ \begin{array}{c} 1 \\ 1
     \end{array}\right],  
   \ket{-} = \frac{1}{\sqrt{2}} \left[ \begin{array}{c} 1 \\ -1
     \end{array}\right].  
\]

It is easy to see that a linear operator is unitary if it preserves the
inner product and hence the norm.  Thus unitaries can be viewed as maps
from quantum states to quantum states.

Some particularly useful unitaries are the \emph{Pauli operators} given by
the following matrices in the canonical basis of $\ctwo$: 
\AR{
X=\MA{0&1\\1&0},&
Y=\MA{0&{-i}\\i&0},&
Z=\MA{1&0\\0&{-1}}
}
We note that all these operators are involutive, self-adjoint, and
therefore unitaries.  
All these matrices have determinant $=-1$.
We will not discuss the geometric significance of
these operators here; their real importance in quantum mechanics comes from
the fact that they can be used to describe rotations, thus they are usually
called the ``Pauli spin matrices'' by physicists.

\newcommand{\cv}[2]{\mbox{$\begin{pmatrix} #1\\#2\end{pmatrix}$}}
\newcommand{\icv}[2]
{\mbox{$\bigl(\begin{smallmatrix} #1\\#2\end{smallmatrix}\bigr)$}}

Some basic algebra of these matrices are given below.
First they all square to the identity.
\[X^2 = Y^2 = Z^2 = I.\]
The Pauli operators do not commute.  We use the notation $[A,B]$ for
$AB-BA$, the \emph{commutator} of $A$ and $B$.  The commutator measures the
extent to which two operators fail to commute: it is customary to
present the algebra of operators using it.  One also uses the symbol
$\{A,B\}$ to stand for $AB + BA$: it is called the \emph{anti-commutator}.
For the Pauli operators we have the following commutators and
anti-commutators :
\begin{tabbing}
aaaa\=aaaaaaaaaaaaaaaa\=aaaaaaaaaaaaaaaa\=aaaaaaaaaaaaaaaa\=a\kill\\
\> $XY = iZ$ \> $YX = -iZ$
\> $[X,Y] = 2iZ$\> $\{X,Y\}=0$\\
\> $ZX = iY$ \> $XZ = -iY$
\> $[Z,X] = 2iY$\> $\{Z,X\}=0$\\
\> $YZ = iX$ \> $ZY = -iX$
\> $[Y,Z] = 2iX$ \> $\{Y,Z\}=0$
\end{tabbing}

\DE
Define the \emph{Pauli group}, $\mathbf{P}_n$, as the group consisting of
tensor products of I, X, Y, and Z on n qubits, with an overall phase of
$\pm 1$ or $\pm i$. 
\ED

Given a group $G$ the operation $x \mapsto g^{-1}xg$ is called
\emph{conjugation} by $g$.  These conjugations give the effect of switching
operators around.  If $G$ is a group and $H$ is a subgroup of $G$ then the
\emph{normalizer} of $H$ is another subgroup of $G$, say $K$, with the
property that for all $h\in H,k\in K$ we have $k^{-1}hk$ in $H$.

The effect of conjugating measurements and other corrections by Pauli
operators is a key part of the rewrite rules described in the main text.
They can be verified using the algebra given here.

A very important related group is called the Clifford group.  
\DE The
\emph{Clifford group}, $\mathbf{C}_n$, is the group of unitary operators
that leave the Pauli group invariant under conjugation, i.e.\ it is the
normalizer of the Pauli group viewed as a subgroup of the unitary group.
\ED

 The Clifford group on n qubits can be generated by the Hadamard transform,
 the controlled-$X$ ($CNOT$) or controlled-$Z$ ($\ctR Z$), and the single-qubit phase rotation: 
 \AR{
H=\ost\MA{1&1\\1&-1},&
CNOT=\MA{1&0&0&0\\ 0&1&0&0\\0&0&0&1\\0&0&1&0},&
\ctR Z=\MA{1&0&0&0\\ 0&1&0&0\\0&0&1&0\\0&0&0&-1},&
P=\MA{1&0\\0&i}
}
  
The importance of the Clifford group for quantum computation is that a
computation consisting of only Clifford operations on the computational
basis followed by final Pauli measurements can be efficiently simulated by
a classical computer, this is the Gottesman-Knill theorem
\cite{Gottesman97,NC00}.

\subsection{Density Matrices}

In order to capture partial information about quantum
systems one uses
\emph{density matrices}.  
Before we describe density
matrices we review some linear algebra in the bra-ket
notation.  Given a ket $\ket{\psi}$ the
notation $\proj{\psi}$ denotes the projection operator onto
the one dimensional subspace spanned by $\ket{\psi}$.  To
verify this note that
\[ (\proj{\psi})(\ket{\psi} = \ket{\psi}\norm{\psi} =
\ket{\psi}\] and
\[ (\proj{\psi})(\ket{\phi}) = \ket{\psi}\braket{\psi}{\phi}
=
\braket{\psi}{\phi}\ket{\psi}.  \]  If $\ket{\psi_i}$ is an
orthonormal basis for $\hilb$ the identity matrix is written
$\sum_i\proj{\psi_i}$.  If $Q$ is a linear operator with
eigenvalues $q_i$ and eigenvectors $\ket{q_i}$, which form
an orthonormal basis for $\hilb$, we can represent $Q$ as
$\sum_i q_i \proj{q_i}$.  To see this, let 
$\ket{\psi} = \sum_i c_i \ket{q_i}$ then
\[ Q\ket{\psi} = \sum_i c_i Q\ket{q_i} = \sum_i c_i
q_i\ket{q_i} \] now using our representation for $Q$ we
calculate 
\[ Q\ket{\psi} = \sum_i q_i \proj{q_i}(\ket{\psi}) = 
\sum_{i,j} c_j q_i \ket{q_i}\braket{q_i}{q_j} = \sum_i c_i
q_i \ket{q_i}.
\] 

This is a version of the spectral theorem that we mentioned in the first
subsection of this appendix.

A state (i.e.  a ray in $\hilb$) is called a \emph{pure} state.  If $a$ and
$b$ are distinct eigenvalues of some observable $A$ with corresponding
eigenvectors $\ket{a}$ and $\ket{b}$ it is perfectly possible to prepare a
state of the form $\frac{1}{\sqrt{2}}(\ket{a}+\ket{b})$.  A measurement of
$A$ on such a state will yield either $a$ or $b$ each with probability
$\frac{1}{2}$.  However, it is also possible that a mixture is prepared.
That is to say instead of a quantum superposition a classical stochastic
mixture is prepared.  In order to describe these we will use density
matrices.

For a system in a pure state $\ket{\psi}$, the density
matrix is just the projection operator $\proj{\psi}$.  If we
have an observable $Q$ with eigenvalues $q_i$ -- assumed
nondegenerate for simplicity -- then we can expand
$\ket{\psi}$ in terms of the eigenvectors by
\[ \ket{\psi} = \sum_i c_i\ket{q_i}.\] Now the probability
of observing $q_i$ when measuring $Q$ in the state
$\ket{\psi}$ is $|\braket{q_i}{\psi}|^2$.  Recalling that
the identity is given by $I = \sum_j \proj{q_j}$ we get that
\[ Prob(q_i,\ket{\psi}) =
\sum_j\braket{q_i}{\psi}\braket{\psi}{q_j}\braket{q_j}{q_i}
\] which after rearranging and using the definition of trace
of an operator yields
\[ Tr((\proj{q_i})(\proj{\psi})).\] If as is typical we
write $\rho_{\psi}$ for the density matrix and $P_i$ for the
projection operator onto the subspace spanned by the
eigenvector
$\ket{q_i}$ we get 
\[ Prob(q_i,\ket{\psi}) = Tr (P_i\rho).\] It is an easy
calculation to show that the expectation value for $Q$ in the
state $\ket{\psi}$ is $Tr(Q\rho)$.

What if the state is not known completely?  Suppose that we
only know that a system is one of several possible states
$\ket{\psi_1},\ldots,\ket{\psi_k}$ with probabilities
$p_1,\ldots,p_k$ respectively.  We define the density matrix
for such a state to be
\[ \rho = \sum_{i=1}^k p_i\proj{\psi_i}.\] The same formulas
for the probability of observing a value $q_i$ , i.e.\ 
$Tr(P_i\rho)$ and for the expectation value of $Q$, i.e.\
$Tr(Q\rho)$ apply.  One can check directly that a density
matrix has the following two properties.
\begin{prop} An operator $\rho$ on $\hilb$ is a
\emph{density matrix} if and only if 
\begin{itemize}
\item $\rho$ has trace $1$ and
\item $\rho$ is a positive operator, which means that it has only positive
  eigenvalues or, equivalently, that for any $x\in\hilb$ we have
  $\pair{x}{\rho x}\geq 0$.
\end{itemize} Furthermore, if $\rho$ is a density operator,
$Tr(\rho^2)\leq 1$ with equality if and only if $\rho$ is a
pure state (i.e.\ a projection operator).
\end{prop}

Suppose that we have a density matrix $\rho$ describing a pure state of an
$n+m$ dimensional system.  Now suppose that an observer can only see the
first $n$ dimensions.  The density matrix $\xi$ describing what he can see
is contained by taking the partial trace over the $m$ dimensions that the
observer cannot see.  Doing this gives, in general, a nonpure state.
Similarly a complementary observer who sees only the $m$ dimensions would
construct her density matrix $\sigma$ by taking the appropriate partial
trace.  Taking these traces loses information; in fact, one cannot
reconstruct $\rho$ even from both $\xi$ and $\sigma$.  Certainly the tensor
product of $\xi$ and $\sigma$ does not give back $\rho$.  This is due to
the loss of the cross-correlation information that was encoded in $\rho$
but is not represented in either $\xi$ or $\sigma$.

The axioms of quantum mechanics are easily stated in the language of
density matrices.  For example, if evolution from time $t_1$ to time $t_2$
is described by the unitary transformation $U$ and $\rho$ is the density
matrix for time $t_1$, then the evolved density matrix $\rho'$ for time
$t_2$ is given by the formula $\rho'=U\rho U^\dagger$.  Similarly, one can
describe measurements represented by projective operators in terms of
density matrices \cite{NC00,Preskill98}.  Thus if a projector $P$ acts on a
state $\ket{\psi}$ then the result is $P\ket{\psi}$; the resulting
transofrmation of density matrices is $\ketbra{\psi}\mapsto
P\ketbra{\psi}P$.  For a general density matrix $\rho$ we have $\rho\mapsto
P\rho P$, note that since $P$ is self-adjoint we do not have to write
$P^{\dagger}$.

\subsection{Operations on Density matrices}
What are the legitimate ``physical'' transformations on density matrices?  
Density matrices are positive operators and they have trace either equal to
$1$ if we insist on normalizing them or bounded by $1$.  These properties
muct be preserved by any transformations on them.

We need first to define what it means for a vector to be positive.
Any vector space $V$ can be equipped with a notion of positivity.  
\DE
A subset
$C$ of $V$ is called a \emph{cone} if
\begin{itemize}
\item $x\in C$ implies that for any \emph{positive} $\alpha$, $\alpha x\in C$,
\item $x,y \in C$ implies that $x+y \in C$ and
\item $x$ and $-x$ both in $C$ means that $x=0$.
\end{itemize}
We can define $x \geq 0$ to mean $x\in C$ and $x \geq y$ to mean $x-y\in
C$.
\ED

\DE
An \emph{ordered} vector space is just a vector space equipped with a
cone.  
\ED

It is easy to check the following explicitly.
\PRO
The collection of positive operators in the vector space of linear
operators forms a cone. 
\ORP

Now we can say what it means for a map to be a positive map.
\DE
Abstractly, $L: (V,\leq_V)\to (W,\leq_W)$ is a \emph{positive map} if
\[\forall v\in V.\;\; v\geq_V 0 \Rightarrow L(v)\geq_W 0.\]
\ED
It is important to not confuse ``positive maps'' and ``positive operators.''  

If we are transforming states (density matrices) then the legitimate
transformations obviously take density matrices to density matrices.  They
have to be \emph{positive maps} considered as maps between the appropriate
ordered vector spaces.  The appropriate ordered vector spaces are the
vector spaces of linear operators on $\hilb$ the Hilbert space of pure states.

Unfortunately the tensor product of two positive maps is not positive in
general.  We really want this!  If one can perform transformation $T_1$ on
density matrix $\rho_1$ and transformation $T_2$ on density matrix $\rho_2$
then it should be possible to regard $\rho_1\otimes\rho_2$ as a composite
system and 
carry out $T_1\otimes T_2$ on this system.  We certainly want this if, say,
$T_2$ is the identity.  But even when $T_2$ is the identity this may fail;
the usual example is the transposition map, see, for example~\cite{NC00}.

The remedy is to require the appropriate condition by fiat.
\DE
A \emph{completely positive map} $K$ is a positive map such that for
every identity map $I_n: \mathbb{C}^n\to\mathbb{C}^n$ the tensor product
$K\otimes I_n$ is positive.
\ED
It is not hard to show that the tensor of completely positive maps is
always a completely positive map.  This condition satisfies one of the
requirements.  We can insist that they preserve the bound on the trace to
satisfy the other requirement as well.  However we would like an explicit
way of recognizing this.

\newcommand{\cB}{\mathcal{B}}
\newcommand{\cE}{\mathcal{E}}

The important result in this regard is the Kraus representation
theorem~\cite{Choi75}.

\TH[Kraus]
The general form for a completely positive map
$\cE:\cB(\hilb_1)\to\cB(\hilb_2)$ is 
\[ \cE(\rho) = \sum_m A_{m}\rho\adj{A_{m}} \]
where the $A_m:\hilb_1\to\hilb_2$.  
\HT
Here $\mathcal{B}(\hilb)$ is the Banach space of bounded linear operators
on $\hilb$.  
If, in addition, we require that the trace
of $\cE(\rho)\leq 1$ then the $A_m$ will satisfy
\[ \sum_m \adj{A_m}A_m \leq I.\]

The following term is common in the quantum computation literature.
\DE
A \emph{superoperator} $T$ is a linear map from $\bhil V$ to $\bhil U$ that
is completely positive and trace preserving.
\ED

\end{document}